\documentclass[useAMS,usenatbib]{mn2e}
\usepackage{graphicx,times}
\usepackage{amssymb,amsmath,color,soul}
\usepackage{deluxetable,natbib}
\usepackage[colorlinks = true,linkcolor = blue,urlcolor  = blue,citecolor = blue,anchorcolor = blue]{hyperref}

\DeclareRobustCommand{\ion}[2]{%
\relax\ifmmode
\ifx\testbx\f@series
{\mathbf{#1\,\mathsc{#2}}}\else
{\mathrm{#1\,\mathsc{#2}}}\fi
\else\textup{#1\,{\mdseries\textsc{#2}}}%
\fi}

\newcommand{\OI}{[\ion{O}{i}]} 
\newcommand{\OIII}{[\ion{O}{iii}]} 
\newcommand{\NII}{[\ion{N}{ii}]}
\newcommand{\SII}{[\ion{S}{ii}]}  
\newcommand{\CII}{\rm[\ion{C}{ii}]~$\lambda158\mu\rm m$}

\newcommand{\source} {SDSS J090005.05+000446.7}
\newcommand{\shortsource} {SDSS J0900}
\newcommand{\lzifu} {{\scshape lzifu}}
\usepackage{url}
\usepackage{subfig}
\usepackage{float}

\title[SAMI: Shocks and Outflows in a normal star-forming galaxy]{The SAMI Galaxy Survey: Shocks and Outflows in a normal star-forming galaxy}

\author[Ho et al.]
  {I-Ting Ho$^1$, Lisa J. Kewley$^{1,2}$, Michael A. Dopita$^{2,3}$, Anne M. Medling$^2$, 
  \newauthor J. T. Allen$^{4,5}$, Joss Bland-Hawthorn$^4$, Jessica V. Bloom$^{4,5}$, Julia J. Bryant$^{4,5,6}$, 
  \newauthor Scott M. Croom$^{4,5}$, L. M. R. Fogarty$^{4,5}$, Michael Goodwin$^6$, Andy W. Green$^6$, 
  \newauthor  Iraklis S. Konstantopoulos$^{6,5}$, Jon S. Lawrence$^6$, \'A.R. L\'opez-S\'anchez$^{6,7}$, 
  \newauthor Matt S. Owers$^6$, Samuel Richards$^{4,5,6}$, and Rob Sharp$^2$\\ 
  $^1$Institute for Astronomy, University of Hawaii, 2680 Woodlawn Drive, Honolulu, HI 96822, USA \\
  $^2$Research School of Astronomy and Astrophysics, Australian National University, Cotter Rd., Weston ACT 2611, Australia\\
  $^3$Astronomy Department, King Abdulaziz University, P.O. Box 80203, Jeddah, Saudi Arabia\\
  $^4$Sydney Institute for Astronomy (SIfA), School of Physics, University of Sydney, NSW 2006, Australia\\
  $^5$ARC Centre of Excellence for All-sky Astrophysics (CAASTRO)\\
  $^6$Australian Astronomical Observatory, PO Box 915, North Ryde, NSW 1670, Australia\\
  $^7$Department of Physics and Astronomy, Macquarie University, NSW 2109, Australia}

\begin{document}
\date{Accepted ??. Received ??; in original form ??}

\label{firstpage}

\maketitle

\begin{abstract}
We demonstrate the feasibility and potential of using large integral field spectroscopic surveys to investigate the prevalence of galactic-scale outflows in the local Universe. Using integral field data from the Sydney-AAO Multi-object Integral field spectrograph (SAMI) and the Wide Field Spectrograph, we study the nature of an isolated disk galaxy, \source\ ($z = 0.05386$). In the integral field datasets, the galaxy presents skewed line profiles changing with position in the galaxy. The skewed line profiles are caused by different kinematic components overlapping in the line-of-sight direction. We perform spectral decomposition to separate the line profiles in each spatial pixel as combinations of (1) a narrow kinematic component consistent with HII regions, (2) a broad kinematic component consistent with shock excitation, and (3) an intermediate component consistent with shock excitation and photoionisation mixing. 
The three kinematic components have distinctly different velocity fields, velocity dispersions, line ratios, and electron densities. We model the line ratios, velocity dispersions, and electron densities with our {\scshape mappings~iv} shock and photoionisation models, and we reach remarkable agreement between the data and the models. The models demonstrate that the different emission line properties are caused by major galactic outflows that introduce shock excitation in addition to photoionisation by star-forming activities. Interstellar shocks embedded in the outflows shock-excite and compress the gas, causing the elevated line ratios, velocity dispersions, and electron densities observed in the broad kinematic component. We argue from energy considerations that, with the lack of a powerful active galactic nucleus, the outflows are likely to be driven by starburst activities.  
Our results set a benchmark of the type of analysis that can be achieved by the SAMI Galaxy Survey on large numbers of galaxies. 
\end{abstract}


\section{Introduction}

Galactic-scale outflows, driven by stellar feedback and/or active galactic nuclei (AGNs), play a critical role in galaxy formation and evolution. Studies of galactic-scale outflows in the local Universe reveal that outflows can carry large amounts of energy and mass and can substantially affect star formation and the chemical evolution of galaxies. In galaxy mergers, powerful galactic-outflows driven by AGN feedback and/or intense starbursts can quench star-formation by removing the necessary gas reservoirs (\citealt*{Di-Matteo:2005qy}; \citealt{Hopkins:2006uq,Diamond-Stanic:2012kx}). Some evidence suggests that the outflowing gas is chemically enriched \citep*{Martin:2002lr}; outflows could therefore alter global chemical properties of galaxies and be responsible for shaping the mass-metallicity relation \citep{Tremonti:2004fk}. The metal-enriched outflows are very likely to supply the vast amount of heavy elements discovered in the circumgalactic medium \citep{Tumlinson:2011lr,Werk:2013fk}. The importance of outflows is also emphasized by numerical simulations, which show that many observed galaxy properties, such as stellar-mass functions, cosmic star-formation rates, and mass-metallicity relations, cannot be reproduced without including feedback to prevent over-accretion of the gas, and to remove metals (e.g., \citealt{Oppenheimer:2010lr}; \citealt*{Dave:2011qy,Dave:2011fk,Hopkins:2012uq}).

In-depth studies of starburst-driven outflows have provided great detail of their structures, kinematics, mass outflow rates, and energetics (see \citealt*{Veilleux:2005qy} for a review). In classical nearby systems such as M82, NGC253, NGC1482, NGC3079 and many others, conical bipolar structures extending up to a few kiloparsecs  beyond stellar disks are commonly seen (e.g., \citealt{Bland:1988yq}; \citealt*{Westmoquette:2011uq}; \citealt{Veilleux:2002fk,Cecil:2001uq}). The conical structures are comprised predominately of warm and hot gas emitting in X-ray and optical line emission (e.g., \citealt{Strickland:2000rt}; \citealt*{Cecil:2002vn}); these line-emitting structures are often limb-brightened. Entrained cold molecular gas has also been observed in the radio wavelengths (e.g., \citealt*{Walter:2002fr}; \citealt{Bolatto:2013qy}). The bipolar structures are open-ended and the gas kinematics often show velocity increases with radius, reaching typically a few hundred $\rm km~s^{-1}$ \citep[e.g.,][]{Shopbell:1998yq,Westmoquette:2011uq}.

Whilst detailed studies in nearby systems yield insight into their nature, the prevalence of galactic outflows is still largely unknown. Starburst-driven outflows are found to be ubiquitous in galaxies with high star formation rate surface densities (e.g., $\rm\Sigma_* \geq 0.1~M_\odot~yr^{-1}~kpc^{-2}$, \citealt{Heckman:2002fk}). Outflows are common in local galaxies with extreme star-forming activities \citep*[e.g., $z<0.5$;][]{Rupke:2005fv,rupke:2005b}, and are ubiquitous at high redshifts where the cosmic star formation rates are high \citep[e.g., $z>1.4$;][]{Weiner:2009lr,Steidel:2010fk}. Whether typical normal star-forming galaxies with star formation rate surface densities below the empirical threshold can launch outflows requires further investigation (\citealt*{Murray:2011fj}; \citealt{Rubin:2014ly}). 

In the local Universe, probing the prevalence of outflows depends on observing large numbers of galaxies to look for outflow signatures. This is most easily achieved by searching for extra-planar optical emission or blue-shifted absorption lines; however, the current sample sizes remain small. Extra-planar line emission has been studied in sample sizes of few tens galaxies using slit spectroscopy and narrow-band imaging (e.g., \citealt*{Heckman:1990lr}; \citealt{Veilleux:2003fr,Veilleux:2005qy}). Studies of blue-shifted absorption by neutral gas entrained in outflows have reached sample sizes on the order of tens to a hundred and provided constraints on the mass, momentum, and energy of outflows \citep[][and references therein]{Heckman:2000lr,Rupke:2005fv,rupke:2005b,Rubin:2014ly}. To probe absorption features, higher signal-to-noise ratios (S/Ns) are required, so spatially binning the data is usually adopted. This binning directly results in the difficulties in recovering geometric information. Stacking spectra of galaxies of similar properties is sometimes necessary to detect outflow signatures, at the expense of probing mean outflow properties of pre-defined populations \citep[e.g.,][]{Chen:2010qy}. 

Integral field spectroscopy (IFS) is an alternative (and ideal) tool for investigating galactic outflows. With IFS observations of 10 nearby galaxies, \citet{Sharp:2010qy} demonstrate that the structures, kinematics, and excitation of outflows can be studied in unprecedented detail, difficult to achieve with slit or fiber spectroscopy or narrow-band imaging. Studies with IFS in local luminous and ultra-luminous infrared galaxies (LIRGs and ULIRGs) show that outflows are usually associated with interstellar shocks that excite optical line emission with enhanced line ratios (\citealt{Rich:2010yq}; \citealt*{Rich:2011kx}; see also \citealt{Soto:2012fk} and \citealt{Soto:2012kx}).

The technological advance of IFS has only recently made studying large samples of objects with spatially resolved spectroscopy possible. Surveys with IFS in the local Universe ($z\lesssim0.01$) such as the SAURON survey \citep{Bacon:2001rt} and its extension the $\rm ATLAS^{3D}$ survey \citep{Cappellari:2011ys}, the PINGS survey \citep{rosales-ortega:2010}, the Calar Alto Legacy Integral Field spectroscopy Area survey \citep[CALIFA;][]{Sanchez:2012fj}, and the VENGA survey \citep{Blanc:2013kx} have built up samples on the order of tens to hundreds of galaxies using single integral field units (IFUs). These surveys have delivered remarkable detail through 3-dimensional datasets across different types of galaxies. Surveys with IFS aiming for slightly higher redshifts ($z\sim0.05$) and much larger sample sizes, on the order of thousands to ten thousands, are currently underway by using multiple fiber-bundle IFUs on focal planes to significantly increase the sampling speed. The on-going SAMI Galaxy Survey\footnote{\href{http://sami-survey.org/}{http://sami-survey.org/}} will ultimately deliver a sample of approximately 3,400 galaxies using the 3.9-m Anglo-Australian Telescope \citep{Croom:2012qy,Allen:2015lq}. The SAMI galaxies are selected from the Galaxy And Mass Assembly project \citep[GAMA;][]{Driver:2009ai}, with the addition of 8 clusters, using stellar-mass cut-offs in redshift bins up to $z<0.12$. The details of the sample selection can be found in \citet{Bryant:2015bh}. With its high spectral resolution ($\rm R\approx 4500$) and large sample size, the SAMI Galaxy Survey will provide a unique sample for investigating the prevalence of outflows in the local Universe. The MaNGA\footnote{\href{http://www.sdss3.org/future/manga.php}{http://www.sdss3.org/future/manga.php}} survey, part of the next generation SDSS~IV being conducted on the 2.5-m telescope at Apache Point Observatory, will observe tens of thousands of galaxies.

During the commissioning of the SAMI instrument, \citet{Fogarty:2012kx} presented a serendipitously discovered galactic wind, and demonstrated that the excellent sensitivity of SAMI enables the detection of low surface brightness extra-planar emission from the nearly edge-on disk galaxy. The line emission shows kinematics and line ratios consistent with galactic winds. In this paper, we present complex spectral analysis of a moderately inclined star-forming galaxy observed by SAMI. We spectrally decompose emission lines to separate those from gas of different kinematic signatures overlapping in the line-of-sight direction. Taking full advantage of the high spectral resolution of SAMI, we present a novel spectral decomposition technique that allows us to study the excitation and physical properties of gas of different origins. By modeling the emission line ratios with our state-of-the-art photoionisation and shock models, we show compelling evidence that the galaxy is launching bipolar outflows driven by starbursts. The excellent agreement between the data and shock models set a benchmark for the type of analysis that can be achieved by the SAMI Galaxy Survey on large numbers of galaxies. The robustness of these results are further confirmed by our supplemental observations using the Wide-Field Spectrograph, a wide-field IFS with higher spectral and spatial resolutions than SAMI. 

The paper is structured as follows. In Section~\ref{sec-source}, we present the properties of the galaxy to be analyzed. In Section~\ref{sec-obs}, we describe our observation and data reduction procedures. In Section~\ref{sec-spectral-analysis}, we demonstrate our spectral decomposition techniques. Results are presented in Section~\ref{sec-results}. In Section~\ref{sec-source-nature}, We discuss the nature of the excitation sources for different kinematic components. In Section~\ref{sec-discussion}, we  discuss the driver of the outflows. Finally, the conclusions are given in Section~\ref{sec-conclusions}. 

\begin{figure}
\centering
\includegraphics[width = 8.5cm]{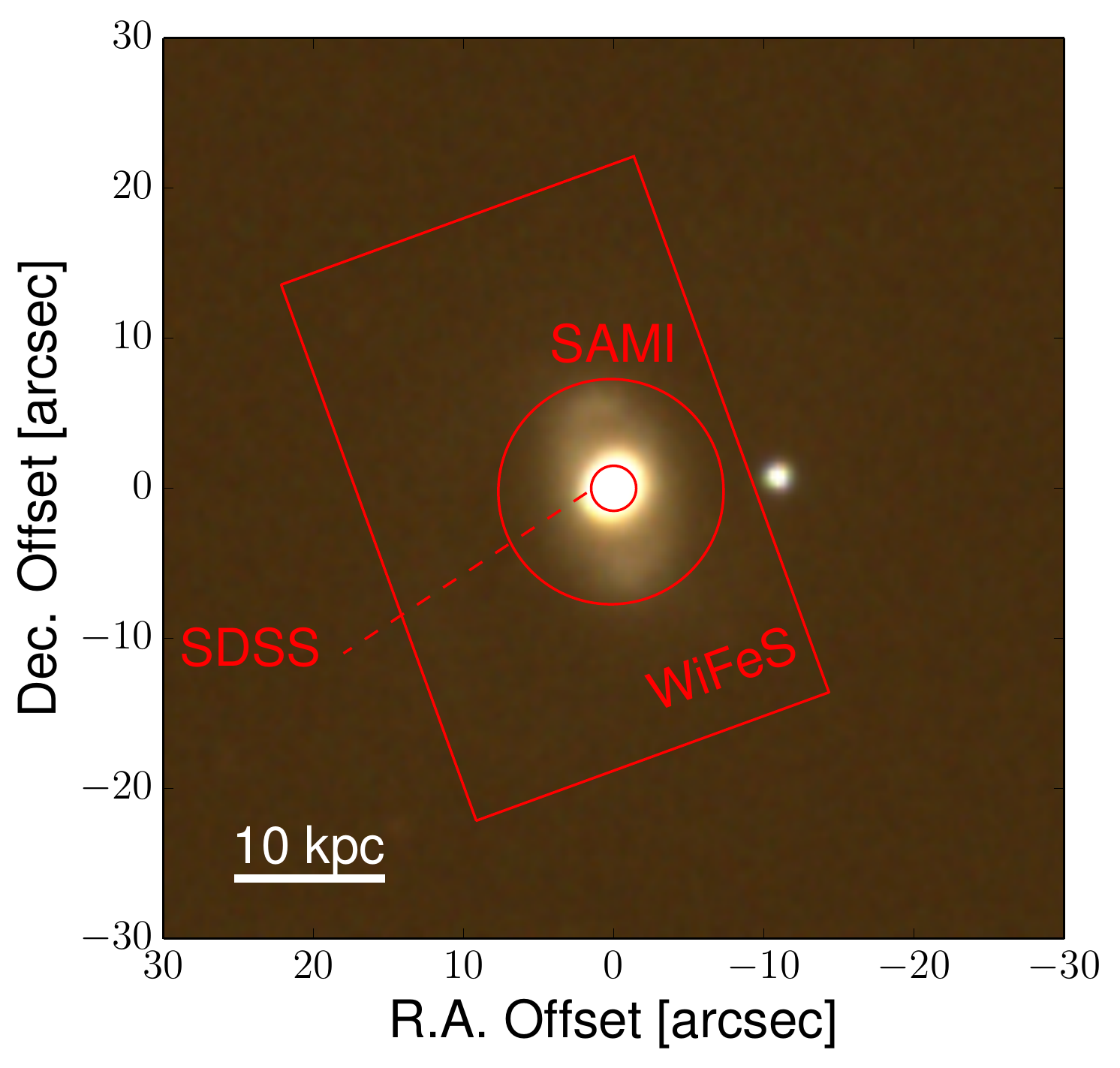}
\caption{SDSS {\it g, r, i} colour composite image over-plotted with footprints of SDSS, SAMI and WiFeS. The SDSS fiber has a diameter of $3\arcsec$. The SAMI hexabundle has a circular field of view of $15\arcsec$ in diameter, and the WiFeS image-slicing IFU has a field of view of $25\arcsec\times38\arcsec$. }\label{pointings}
\end{figure}

\section{\source}\label{sec-source}

\source\ (GAMA ID: 209807; hereafter, \shortsource; Fig.~\ref{pointings}) is an isolated disk galaxy located in the Sloan Digital Sky Survey field \citep[SDSS;][]{York:2000qe}, with a SDSS {\it r}-band magnitude of 15.1, and a logarithmic stellar mass of $\rm 10.8~M_\odot$ \citep{Taylor:2011kx}. The SDSS imaging data do not show any potential companion galaxies within a $1\arcmin$ radius of \shortsource\ (63~kpc at the galaxy redshift); this is consistent with the disk-like morphology and the lack of tidal features of \shortsource.  The inclination angle of \shortsource, as estimated from the axis ratio of 0.75 on the SDSS {\it r}-band image, is approximately $43^\circ$ \citep{Padilla:2008fk}. The SDSS spectrum of \shortsource, taken with optical fibers of 3\arcsec in diameter, captures emission predominately from the nucleus of the galaxy, and contains  only approximately 22\% of the total light. On standard line ratio diagnostic diagrams (the BPT diagrams; \citealt*{Baldwin:1981lr}; \citealt{Veilleux:1987qy}; see below Fig.~\ref{bpt}), the nuclear line ratios fall close ($<0.1$dex) to the maximum theoretical starburst lines parametrized by \citet{Kewley:2001lr}. In the classification scheme by \citet{Kewley:2006lr}, \shortsource\ is formally classified as an ambiguous galaxy. The {[\ion{N}{ii}]~$\lambda$6583}/H$\alpha$ ratio is consistent with composite galaxies, the [\ion{S}{ii}]~$\lambda\lambda$6716,31/H$\alpha$ ratio is consistent with star-forming galaxies, and the [\ion{O}{i}]~$\lambda$6300/H$\alpha$ ratio is consistent with low-ionisation narrow emission-line regions \citep[LINERs;][]{Heckman:1980vn}. 

Through out this paper, we assume the concordance $\Lambda$ cold dark matter cosmology with $ H_0 = 70~\rm km~s^{-1}~Mpc^{-1}$, $\rm \Omega_M = 0.3$ and $\rm \Omega_{\Lambda} = 0.7$, for which the luminosity distance is 240~Mpc and $1\arcsec$ corresponds to 1.05~kpc at the galaxy redshift of $z = 0.05386$. 

\section{Observations and Data Reduction}\label{sec-obs}

\subsection{SAMI}

\shortsource\ was observed on 2013 March 17 using the Sydney-AAO Multi-object Integral field spectrograph \citep[][]{Croom:2012qy} mounted on the 3.9-m Anglo-Australian Telescope. SAMI consists of 13 imaging fiber bundles \citep[`hexabundles';][]{Bland-Hawthorn:2011lr,Bryant:2011qy,Bryant:2014fk} deployable over a 1-degree diameter field of view (FOV). Each hexabundle comprises 61 lightly-fused multimode fibers. Each fiber is $1.6\arcsec$ in diameter, forming  a circular FOV of each hexabundle of $15\arcsec$ in diameter. Despite the high filling factor (75\%) of the hexabundles, $\sim7$ dither positions are required to properly sample the gaps between fibers. For \shortsource, a set of eight 1800-s dithered observations was obtained. Fibers are fed to the flexible AAOmega dual-beam spectrograph \citep{Sharp:2006mz} which can be used at a range of spectral resolutions ($\rm R \equiv \lambda/\delta\lambda \approx 1700\ \mbox{--}\ 13000$) over the optical spectrum ($\rm 3700\ \mbox{--}\ 9500$\AA). 

Standard spectroscopic data reduction procedures were accomplished using the data reduction pipeline {\scshape 2dfdr}\footnote{\href{http://www.aao.gov.au/science/software/2dfdr}{http://www.aao.gov.au/science/software/2dfdr}} \citep{Croom:2004fk}, which performs dark and bias subtraction, long-slit flat fielding, tramline identification, extraction, spectral flat fielding, wavelength calibration, throughput correction and sky subtraction. Flux calibration was performed using observations of the spectrophotometric standard HD111786 taken on the same night. A correction for telluric absorption was made using simultaneous observations of a secondary standard star. Reduced frames using different dithers were reconstructed and resampled on a Cartesian grid. The full procedures for flux calibration, telluric correction and resampling are described in \citep{Sharp:2015ve}, where the data reduction procedures are also reported (see also \citealt{Allen:2015lq}). The final reduced data products comprise two data cubes on spatial grids of $0.5\arcsec\times0.5\arcsec$. The blue data cube covers $\sim3720\ \mbox{--}\ 5850$\AA\ with a spectral sampling of 1.04~\AA\ and spectral resolution of $\rm R\sim1750$ or full-width at half-maximum (FWHM) of $\rm\sim170~km~s^{-1}$. The red data cube covers $\sim6210\ \mbox{--}\ 7370$\AA\ with a spectral sampling of 0.57~\AA\ and spectral resolution of $\rm R\sim4500$ or $\rm FWHM\sim65~km~s^{-1}$. \shortsource\ was observed in relative poor seeing conditions and the point spread function (PSF) has a FWHM of $\sim3.1\arcsec$, as measured from reduced data cubes of a calibration star observed simultaneously with science targets (see below).

\subsection{WiFeS}
To confirm the measurements by SAMI, supplemental observations were also conducted with the Wide Field Spectrograph (WiFeS) on the 2.3-m telescope at Siding Spring Observatory in 2013 November and 2013 December. WiFeS is a dual beam, image-slicing IFU consisting of 25 slitlets, each of which is $38\arcsec$ long and $1\arcsec$ wide, yielding a $25\arcsec\times38\arcsec$ field of view. A more comprehensive description of the instrument can be found in \cite{Dopita:2007kx} and \citet{Dopita:2010yq}. The galaxy was observed with two different spectral settings. For the first setting, we placed B7000 grating on the blue arm and R7000 grating on the red arm. For the second setting, we switched the red arm to I7000 grating in order to capture the {[\ion{S}{ii}]~$\lambda\lambda$6716,31} lines falling outside R7000 spectral coverage. The total effective integration times are 4 hours for the B7000 grating, 2.8 hours for the R7000 grating and 1 hours for the I7000 grating. Typical seeing conditions were $1.5\ \mbox{--}\ 2.5\arcsec$. 

We reduce the data using the custom-built data reduction pipeline {\scshape pywifes} \citep{Childress:2014fr}. Data for each slitlet are reduced following standard spectroscopic data reduction procedures. The reduced data are subsequently reconstructed to form 3-dimensional data cubes. For the sake of convenience, we merged the R7000 and I7000 cubes into a single `red' data cube while keeping the merged B7000 cube as the `blue' data cube. The final reduced cubes have $1\arcsec\times1\arcsec$ in the spatial dimensions. The blue cube covers $\sim4170\ \mbox{--}\ 5550$\AA\ with a spectral resolution of $\rm FWHM\sim40~km~s^{-1}$ ($\rm R = 7000$) and a spectral channel width of 0.35~\AA. The red cube covers $\sim5400\ \mbox{--}\ 8930$\AA\ with a spectral resolution of $\rm FWHM\sim40~km~s^{-1}$ ($\rm R = 7000$) and a spectral channel width of 0.44~\AA. \\

Figure~\ref{pointings} shows the SDSS {\it g, r, i} colour composite image with the FOVs and footprints of SDSS, SAMI and WiFeS over-plotted. As \shortsource\ has an effective radius of 3.5\arcsec, the SAMI aperture covers the galaxy out to approximately 2 effective radii. The WiFeS aperture covers all optical emission visible on the SDSS image.

\section{Spectral Analysis}\label{sec-spectral-analysis}

In this section, we describe our methodology of extracting line fluxes from the data cubes via spectral fitting and statistical tests. The high spectral resolutions of both SAMI and WiFeS can often resolve intrinsic kinematic structures in emission lines, which reveal themselves as skewed line profiles. The profiles can  usually be modeled as superimpositions of Gaussians, assuming that the individual kinematic components have Gaussian line profiles. As we will show, this assumption is validated by the different line ratios (of the different kinematic components) that can be explained by a self-consistent physical picture.  Numerous previous studies have also adopted the methodology of fitting emission lines with multiple Gaussian components, which we refer to here as `spectral decomposition', to gain insight into kinematics and the line ratios of different structural components (e.g., \citealt{Rich:2010yq,Rich:2011kx,Westmoquette:2011uq}; \citealt{Soto:2012fk,Soto:2012kx}). Below, we first describe briefly our spectral fitting pipeline before elaborating on our technique of spectral decomposition.

\subsection{Spectral fitting with \lzifu}

\begin{figure*}
\centering
\includegraphics[width = 16.5 cm]{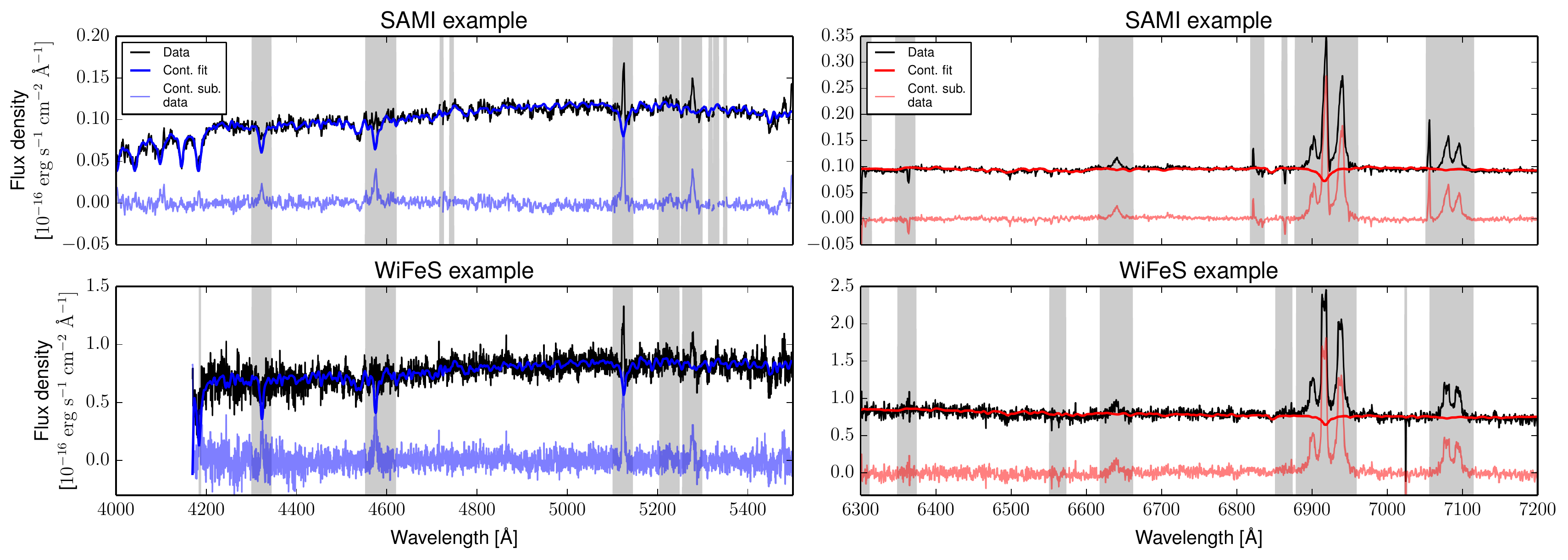}
\caption{Examples of SSP synthesis fitting using {\scshape ppxf} and stellar models by  \citet{Gonzalez-Delgado:2005lr}. The upper panels and lower panels show examples of the SAMI and WiFeS data of \shortsource, respectively. The same two spaxels are also presented in the upper panels of Fig.~\ref{sami_spec} and \ref{wifes_spec} to demonstrate spectral decomposition. The blue (left panels) and the red (right panels) spectra are shown separately. Grey bands mark bad channels, channels around sky lines, or strong emission lines that were not included in the continuum fitting. Details are provided in Section~\ref{sec-spectral-analysis}.  }\label{cont_fit}
\end{figure*}

We adopt our spectral fitting pipeline \lzifu\ to perform spectral fitting and spectral decomposition. A more thorough description of \lzifu\ and its application to the SAMI Galaxy Survey will be presented elsewhere (Ho et al. in prep.). Here, we  provide a brief description. 

\lzifu\ is a pipeline, written in the Interactive Data Language ({\scshape idl}), mainly designed to perform flexible emission line fitting in two-sided IFU data cubes. \lzifu\ first models and subsequently removes the continuum in each spatial pixel (spaxel) by performing simple stellar population (SSP) synthesis. To achieve this, \lzifu\ adopts the penalized pixel-fitting routine developed for the SAURON and $\rm ATLAS^{3D}$ projects \citep[{\scshape ppxf};][]{Cappellari:2004uq}. In this application, we adopt theoretical SSP models, assuming Padova isochrones, of 18 ages\footnote{0.004, 0.006, 0.008, 0.011, 0.016, 0.022, 0.032, 0.045, 0.063, 0.089, 0.126, 0.178, 0.251, 0.355, 0.501, 0.708, 1.000, and 1.413 Gyr.} and solar metallicity from \citet{Gonzalez-Delgado:2005lr}. The primary objective of performing SSP synthesis fitting is to correct for the underlying Balmer absorption caused by the stellar atmospheres of the old stellar population. In Fig.~{\ref{cont_fit}}, we show the SSP synthesis fits of two typical spaxels in the SAMI and WiFeS data. After removing the continuum, \lzifu\ models user-assigned emission lines as Gaussians and performs a bounded value nonlinear least-squares fit using the Levenberg-Marquardt least-squares method implemented in {\scshape idl} \citep[{\scshape mpfit};][]{Markwardt:2009lr}. Users have the options to model each emission line as 1, 2, or 3-component Gaussians describing (potentially) different kinematic components. \lzifu\ automatically establishes reasonable initial guess(es) to proceed with the Levenberg-Marquardt least-squares method. All the user-assigned lines are fitted simultaneously and every kinematic component is constrained to share the same velocity and velocity dispersion. When more than 1 component is fit, \lzifu\ sorts and groups the results based on the velocity dispersions of the different components. The main products delivered by \lzifu\ are two-dimensional (2D) emission line maps useful for emission line studies. Where possible, \lzifu\ simultaneously uses data from the blue and red arms of the spectrographs to fit continuum and emission lines. In this application, we fit with \lzifu\ H$\beta$, {[\ion{O}{iii}]~$\lambda\lambda$4959,5007} \footnote{{[\ion{O}{iii}]~$\lambda$5007}   is fixed to 3 times {[\ion{O}{iii}]~$\lambda$4959} in flux}, {[\ion{O}{i}]~$\lambda$6300}, {[\ion{N}{ii}]~$\lambda\lambda$6548,83} \footnote{{[\ion{N}{ii}]~$\lambda$6583} is fixed to 3 times {[\ion{N}{ii}]~$\lambda$6548} in flux}, H$\alpha$, and {[\ion{S}{ii}]~$\lambda\lambda$6716,31}. Hereafter, we omit the wavelength subscription when appropriate, i.e., \OIII $\equiv$ {[\ion{O}{iii}]~$\lambda$5007}, \NII $\equiv$ {[\ion{N}{ii}]~$\lambda$6583}, \OI $\equiv$ {[\ion{O}{i}]~$\lambda$6300}, and \SII $\equiv$ {[\ion{S}{ii}]~$\lambda\lambda$6716,31}.

\subsection{Spectral decomposition}

\begin{figure*}
\centering
\includegraphics[width = 16cm]{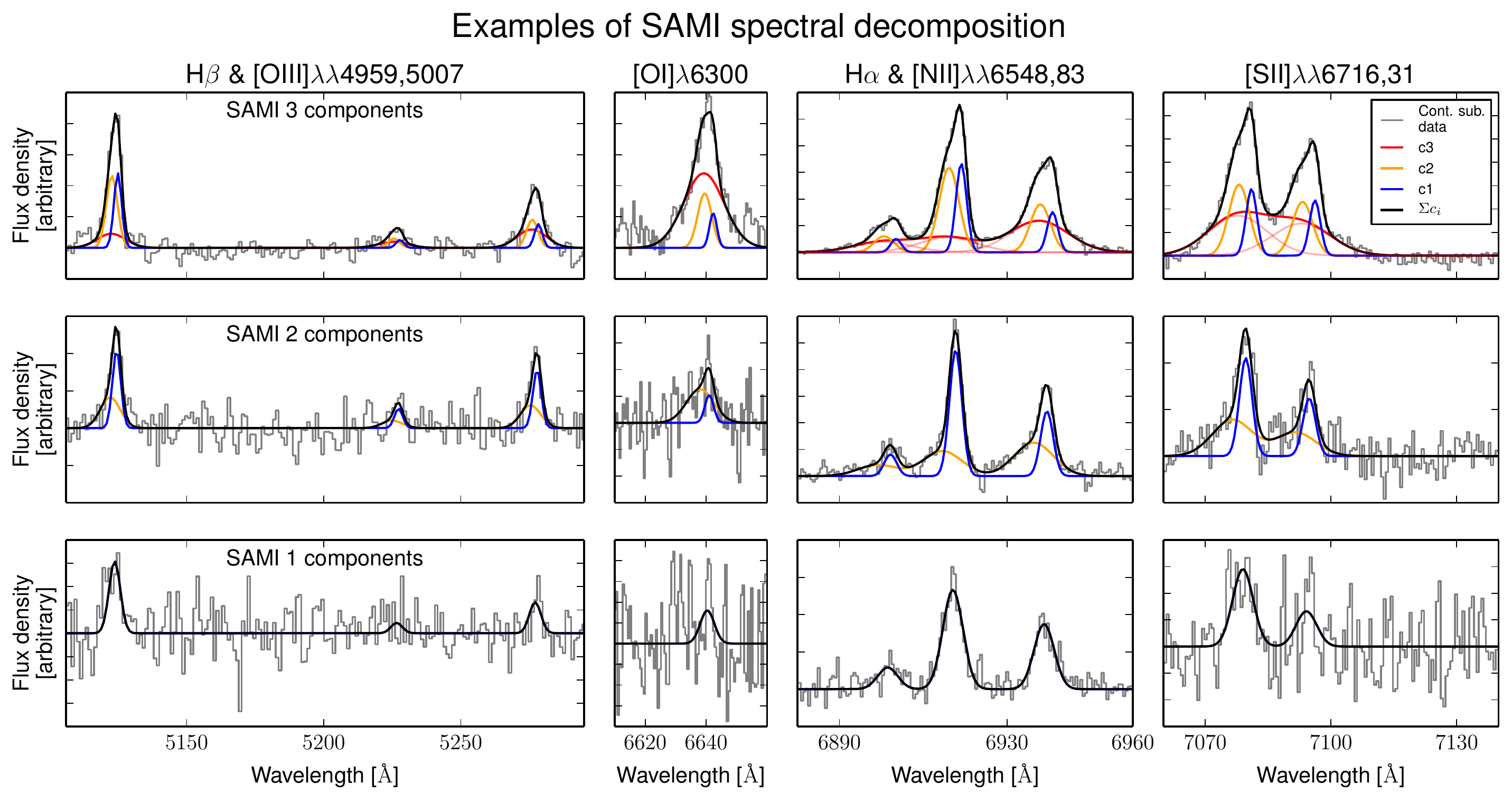}
\caption{Examples of our spectral decomposition using the SAMI data. First, second and third rows are examples of selected spaxels requiring 3-component, 2-component and 1-component fits to describe the spectral profiles, respectively. Each panel is a zoom-in of the spectral ranges comprising key diagnostic emission lines.  Continuum subtracted spectra are shown in grey. Best-fit {\it c3}, {\it c2}, and {\it c1} are shown in red, orange, and blue, respectively. The best-fitting emission line models, i.e. $\sum{c{_i}}$, are shown in black. To avoid confusion in cases where the lines are broad and blended, we also show individual line transitions of {\it c3} as transparent red lines. Details of our spectral analysis technique are described in Section~\ref{sec-spectral-analysis}. }\label{sami_spec}
\end{figure*}

\begin{figure*}
\centering
\includegraphics[width = 16cm]{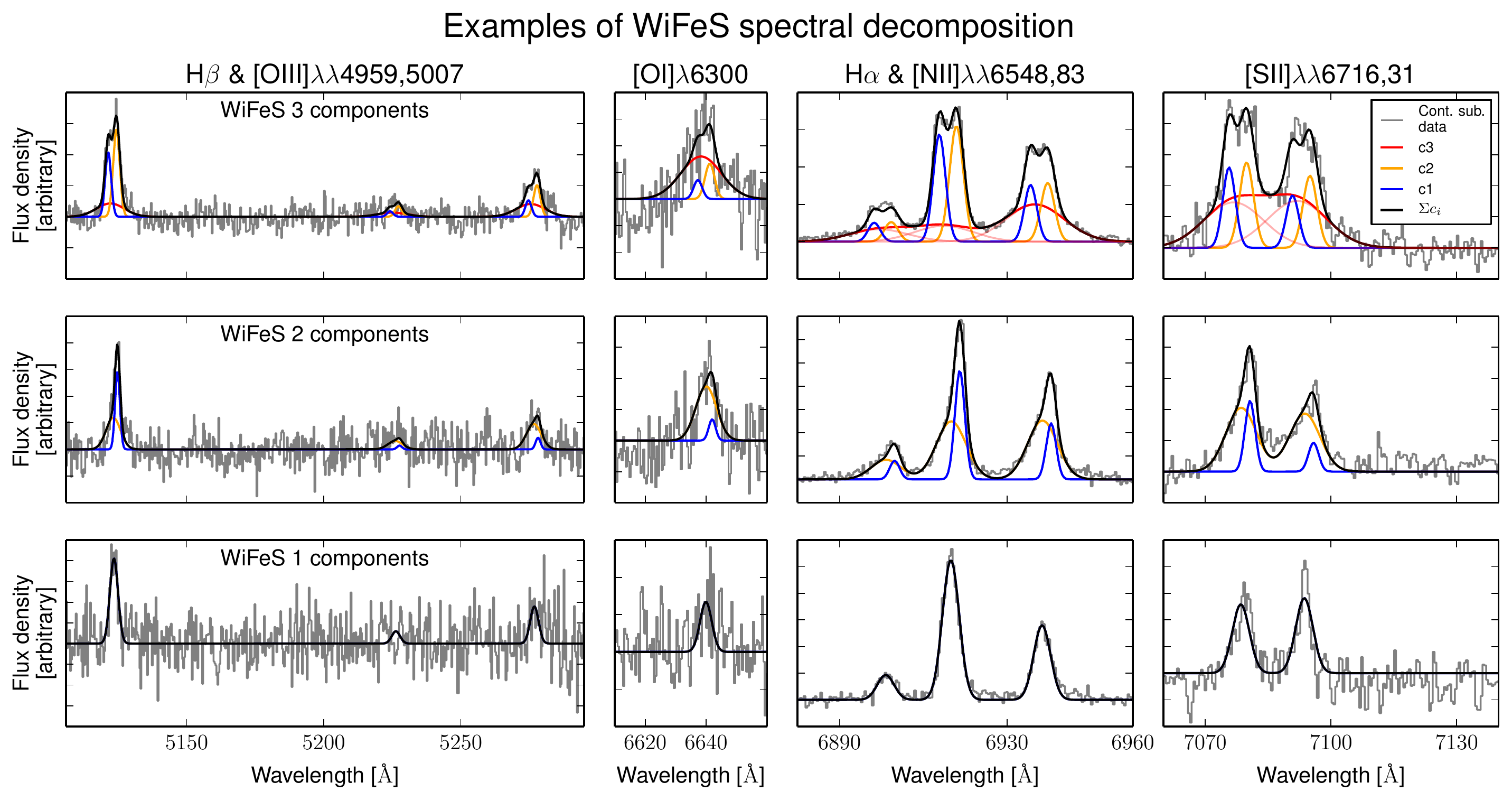}
\caption{Same as Fig.~\ref{sami_spec} but for the WiFeS data.}\label{wifes_spec}
\end{figure*}

How well one can decompose a particular emission line spectrum depends on various factors. One obvious factor is the spectral resolution, which is a gauge of the critical wavelength separation between two different kinematic components (of infinitely small intrinsic line-widths) below which the components cannot be resolved by the instrument. Another factor is the intrinsic velocity dispersions in the kinematic structures. Separating two narrow lines is much easier than separating two broad lines for a given spectral resolution and wavelength separation. In practice, these two factors only serve as rough indicators, as the ability to separate different kinematic components is critically affected by the S/N of the spectrum. When two underlying components have a separation comparable to the spectral resolution, a high S/N spectrum is easier to decompose than a low S/N spectrum. In IFS observations of moderately high spectral resolution ($\rm R\sim4000-7000$; such as SAMI and WiFeS) targeting galaxies with ISM conditions similar to local galaxies, typically the separation between different components is comparable to the spectral resolution, which yields skewed or asymmetric line profiles. For example, $\rm R=4000$ corresponds to a resolving power of  $\rm FWHM\sim75~km~s^{-1}$, which is comparable to typical velocity separation of different kinematic components ($\rm\sim100~km~s^{-1}$) and not significantly smaller than the velocity dispersion typically observed in the ISM (few tens $\rm km~s^{-1}$) and shocked gas ($\rm\lesssim400~km~s^{-1}$). In order to objectively determine whether a spectrum can be decomposed into multiple Gaussian components, some statistical tests taking into account the effect of noise have to be performed.

A common test adopted in many studies is the likelihood ratio test (LRT). While comparing different models, e.g., a {\it n-}component model and a {\it (n+1)-}component model, the (natural) logarithmic maximum likelihood ratio of the two models,
\begin{equation}\label{eq-lrt}
\Lambda = -2 \ln{ {\rm max}(L_n) \over {\rm max}(L_{n+1}) } = \chi^2_n -  \chi^2_{n+1},
\end{equation}
is an objective gauge of how much improvement in maximum likelihood, ${\rm max}(L_n)$ and ${\rm max}(L_{n+1})$, the more sophisticated model can offer. Here, $\chi^2_n$ and $\chi^2_{n+1}$ are $\chi^2$ values of the best fit models with $\nu_n$ and $\nu_{n+1}$ degrees of freedom, respectively. In the case of nested models, where the simpler model (i.e., {\it n-}component model) is a subset of the more sophisticated model (i.e., {\it (n+1)-}component model), and when the numbers of measurements are large (e.g., many spectral channels), $\Lambda$ follows a $\chi^2$ distribution of ($\nu_{n} - \nu_{n+1}$) degree of freedom. The null hypothesis that the {\it n-}component model is superior than the {\it (n+1)-}component model can be tested by comparing the measured $\Lambda$ with the critical $\Lambda$ ($\Lambda_c$) corresponding to a probability {\it p-}value. Representing a statistically significant level, the {\it p-}value is the integration of the $\chi^2$ probability density function from $\Lambda_c$ to infinity.  The {\it p-}value is typically set to 0.05, but in this study we adopt a more conservative value of 0.01. An intuitive way to imagine the LRT is that a more sophisticated {\it (n+1)-}component model  always results in smaller residuals and a lower $\chi^2$ value. The LRT statistically evaluates whether, given the data, the new $\chi^2$ value is small enough to be worth including the extra free parameters to the {\it n}-component model. 

The LRT and a similar statistical {\it F}-test are commonly adopted in many studies performing spectral decomposition \citep[e.g.,][]{Westmoquette:2009a,Westmoquette:2011uq}.  However, \citet{Protassov:2002lr} point out that the use of the LRT and {\it F-}test is not statistically justified in many line-fitting applications. One of the conditions of the LRT and {\it F-}statistics is not strictly satisfied due to the boundary conditions of non-negative line fluxes imposed on the models. The likelihood ratio therefore does not necessary follow the same asymptotic behavior as the $\chi^2$ distribution. Although \citet{Protassov:2002lr} provide alternative statistical tests based on a Bayesian framework and Monte Carlo simulations, the lack of simple standard statistical tests has led some authors to visually classify every spectrum, or combine both the LRT (or {\it F-}test) with visual inspection (e.g., \citealt{Westmoquette:2011uq}; \citealt*{Vogt:2013uq}).

In this work, we perform spectral decomposition with \lzifu, LRT, and visual inspection. We first fit each spaxel with 1, 2, and 3 components with \lzifu, and record the $\chi^2$ of each fit. We then determine which fit best describes a spectrum by calculating the critical likelihood ratio above which the null hypothesis, that the simpler model is more suitable, can be rejected at a significance of 1\%. Each pixel is then assigned to a requirement of either 1, 2, or 3 components and the corresponding fit results are adopted. We record the 2D distribution of  the numbers of components required, and refer it as `component map'. Given the caveat of this statistical test as discussed in \citet{Protassov:2002lr}, we visually inspect randomly selected spaxels and reach the conclusion that LRT gives sensible classification in \shortsource. We also perform simple Monte Carlo simulations to probe the asymptotic behavior of the likelihood ratios. From the statistical aspects, we find that the false classification rates are low ($\lesssim10\%$), and unlikely to impact any of our major conclusions. A more thorough investigation of using LRT and {\it F-}test in spectral decomposition will be presented in Ho et al. (in prep.).

In Fig.~\ref{sami_spec} and \ref{wifes_spec}, we show examples of our spectral decomposition of the SAMI and WiFeS data, respectively. In all cases, the continuum subtracted spectra (grey) are reasonably flat with no evidences of non-zero residuals in line-free channels, indicating that {\scshape ppxf} reliably models the continua. 
 The continuum models of the spaxels shown in the top two row of Fig.~\ref{sami_spec} and \ref{wifes_spec} were presented in Fig.~\ref{cont_fit} for comparison. In the continuum-subtracted data, skewed and non-Gaussian line profiles are obvious in the 3-component and 2-component cases, and our fits reliably reproduce the profiles. In the rest of the paper, we refer to {\it c1} as the component with the smallest velocity dispersion in one particular spaxel, {\it c2} (if a robust fit is obtained according to the LRT) with the middle velocity dispersion, and {\it c3} (if a robust fit is obtained according to the LRT) with the largest velocity dispersion. 

We note that uncertainties in SSP fitting can affect the low amplitude broad kinematic component {\it c3}, particularly in the Balmer lines due to their sensitivity to stellar ages and metallicities. We have performed SSP fitting using different stellar templates and wavelength ranges; we have also compared the emission-line fits with and without including the Balmer lines. We find that our major conclusions are not sensitive to the uncertainties in SSP fitting.

\begin{figure}
\centering
\includegraphics[width = 8.5cm]{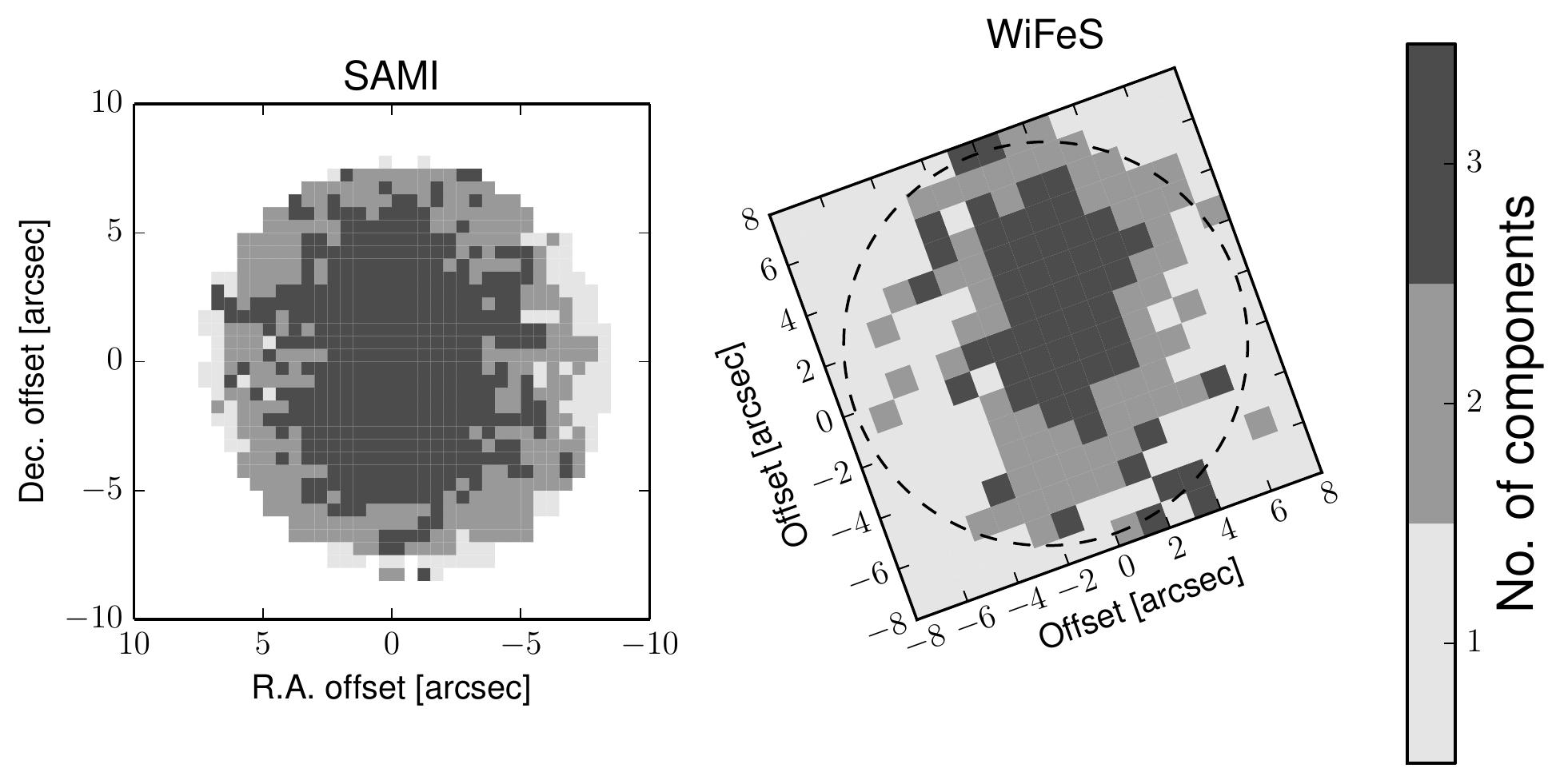}
\caption{Component maps of the SAMI data (left) and the WiFeS data (right) describing the number of components required to model the spectral profiles in each spaxel. The white area in the SAMI map (outside the aperture) contains no data and was not analyzed. The dashed circle in the right panel indicates approximately the SAMI aperture in the WiFeS field. We determine the number of components required by using the statistical LRT, and we also perform visual inspection on selected spectra. Details are provided in Section~\ref{sec-spectral-analysis}.  }\label{comp_map}
\end{figure}

We present in Fig.~\ref{comp_map} the component maps for the SAMI and the WiFeS data. The centre of the galaxy requires a 3-component fit, the number of components required decrease at large radii. Note that whenever S/Ns are sufficiently high ($>3$), the spaxels in the 1-component regions only contribute to {\it c1} to the following analysis, while those in the 3-component regions may contribute to each of {\it c1}, {\it c2}, and {\it c3}. In other words, the narrow component {\it c1} exists in all three zones in Fig.~\ref{comp_map}, but the broader components {\it c2} and {\it c3} are only observed at statistically significant levels in the inner regions.

\section{Results}\label{sec-results}
\subsection{Kinematics of the different components}

\begin{figure}
\centering
\includegraphics[width = 8.5cm]{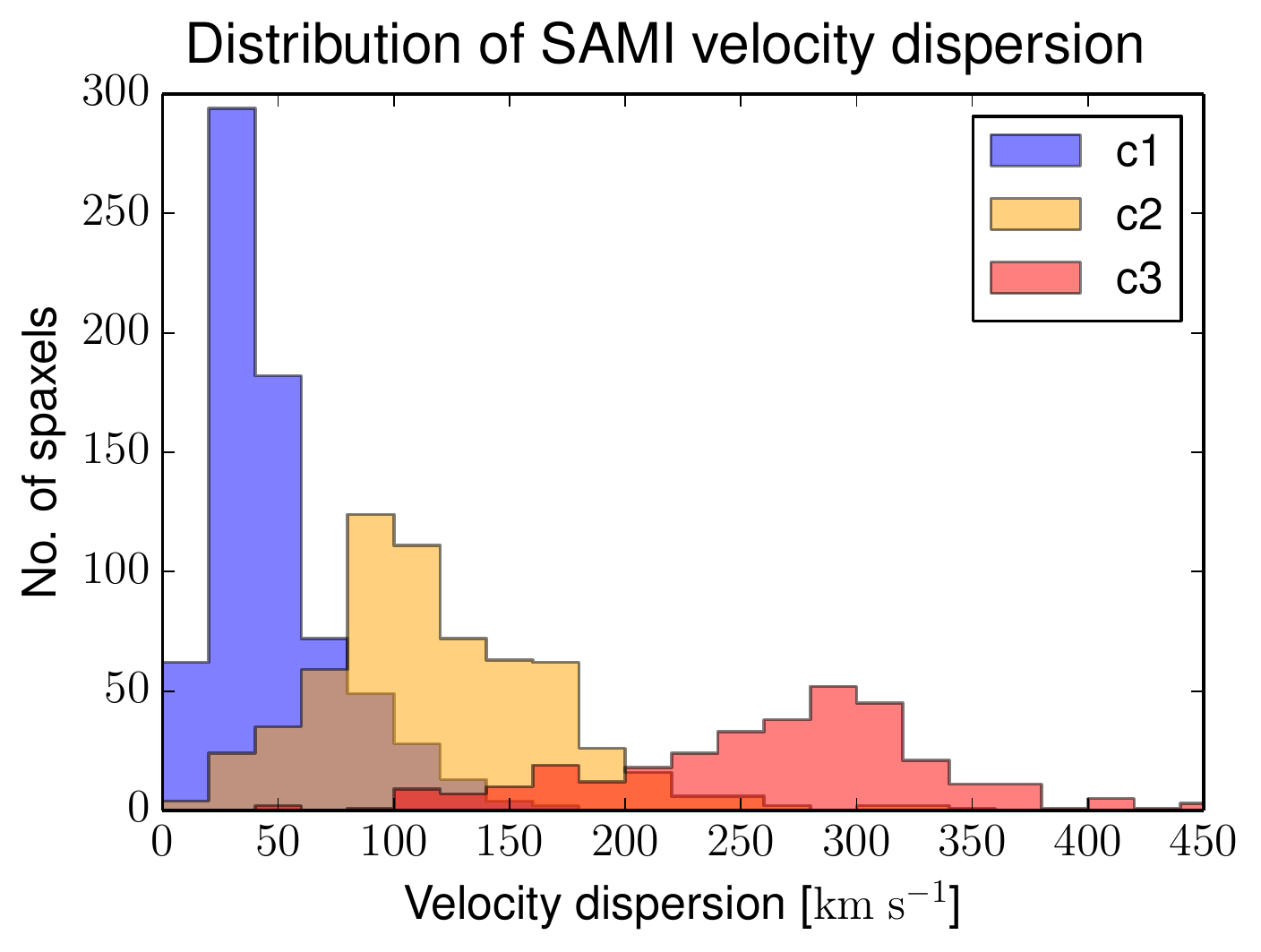}
\caption{The distributions of velocity dispersions of the different components in the SAMI data. The histograms show the distributions of velocity dispersions for every spaxel where H$\alpha$ has $\rm S/N > 3$. It is evident the that different components have distinctly different velocity dispersions, producing a trimodal distribution. }\label{sami_vdisp}
\end{figure}

The distributions of the velocity dispersions of \shortsource\ are shown in Fig.~\ref{sami_vdisp}. We present only spaxels with $\rm S/N > 3 $ for the H$\alpha$ fluxes. It is evident that different components are well separated in velocity dispersion space, producing a trimodal distribution. In particular, the broad kinematic component {\it c3} has distinctly higher velocity dispersions ($\rm\gtrsim150~km~s^{-1}$), with the distribution peaking at approximately $\rm300~km~s^{-1}$. The narrow component {\it c1} has relative low velocity dispersions, with the distribution peaking at approximately $\rm40~km~s^{-1}$.

\begin{figure}
\centering
\includegraphics[width = 8.5cm]{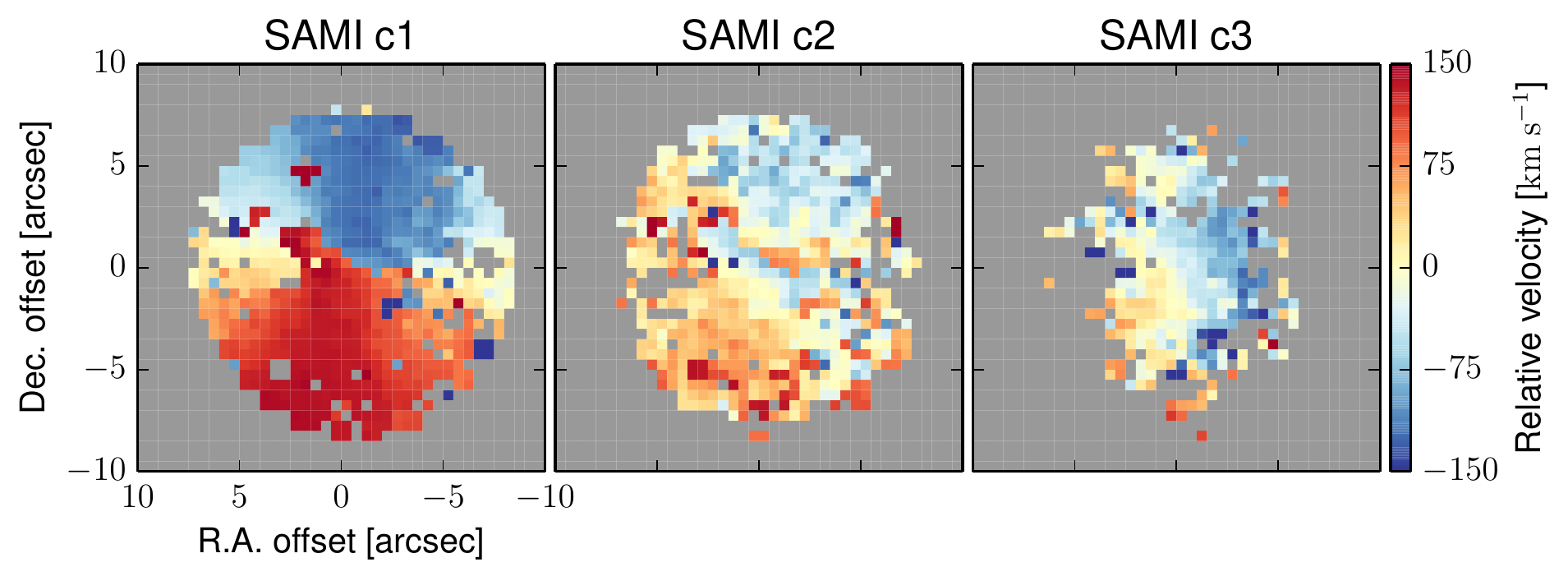}
\caption{Velocity fields of the narrow component {\it c1}, the intermediate component {\it c2}, and the broad component {\it c3} of the SAMI data. Spaxels with H$\alpha$ $\rm S/N < 3$ have been masked out to avoid confusion. The obvious rotation pattern of the narrow component {\it c1}, in the same sense as the disk orientation in Fig.~\ref{pointings}, demonstrates the robustness of our spectral decomposition. The intermediate component {\it c2} and the broad component {\it c3} both show velocity gradients in the north-west to south-east and east to west direction, respectively, with {\it c3} presenting a more prominent and steeper velocity gradient. }\label{sami_v}
\end{figure}

In Fig.~\ref{sami_v}, we present velocity fields of the different components from the SAMI data. Spaxels with low S/Ns ($<3$) in H$\alpha$ fluxes have been masked out. In the left panel, the narrow component {\it c1} traces the rotation field of the disk and shows an obvious rotation pattern in the same sense as the optical disk (Fig.~\ref{pointings}). The regularity of the rotation field further proves the robustness of our spectral decomposition. The intermediate component {\it c2} and the broad component {\it c3} both also show velocity gradients, in the north-west to south-east and east to west direction, respectively, with {\it c3} presenting a more pronounced and steeper velocity gradient. We note that in all three panels, some hot spaxels stand out from the otherwise smooth velocity fields. Some of those spaxels can be removed by raising the S/N cut, but some obviously have bad fit due to the ill-chosen initial guesses in line fitting, or remaining cosmic rays. It is trivial to identify those spaxels and manually refit them with proper initial guesses. We do not perform this exercise because (1) it is impractical to do so for the whole SAMI dataset in the future, and (2) we wish to faithfully present the low but non-zero failure rate of \lzifu. We do take extra caution to ensure that all the conclusions in this paper are independent of these outlying spaxels.

\subsection{Line ratio diagrams and line ratio -- velocity dispersion diagrams}

\begin{figure*}
\centering
\includegraphics[width = 16cm]{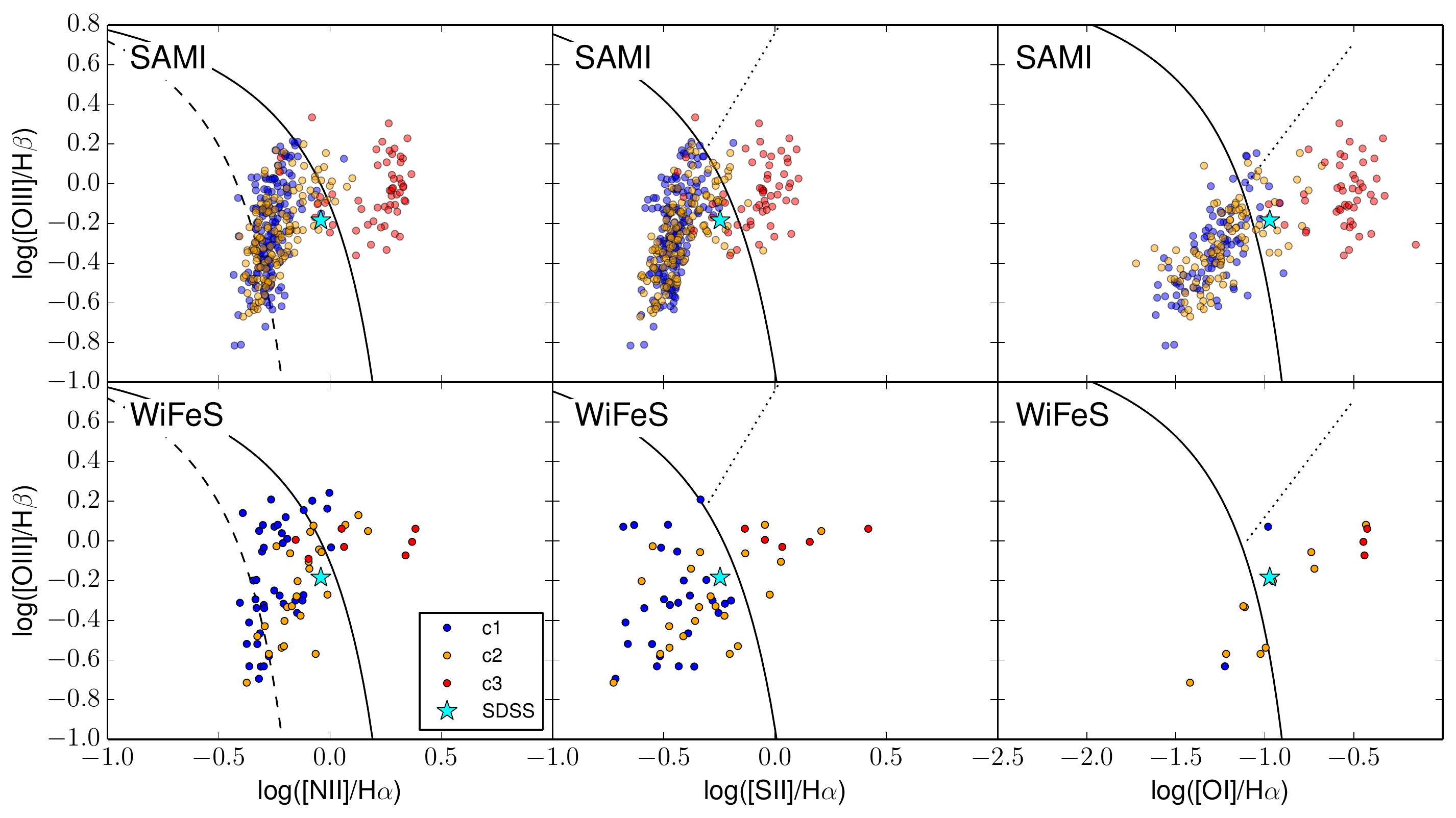}
\caption{
BPT diagrams of the SAMI data (top) and the WiFeS data (bottom). The line ratios of the different components are colour-coded as indicated in the legend. The nuclear line ratios measured by the SDSS spectroscopic survey are shown as stars. The empirical separation lines between starburst galaxies and AGN host galaxies derived from SDSS are shown as dashed lines \citep{Kauffmann:2003vn}. The theoretical extreme starburst lines calculated from photoionisation and radiation transfer models are shown as solid lines \citep{Kewley:2001lr}. The SDSS empirical separation lines between Seyfert galaxies and LINERs are indicated as dotted lines \citep{Kewley:2006lr}. }\label{bpt}
\end{figure*}

\begin{figure*}
\centering
\includegraphics[width = 16cm]{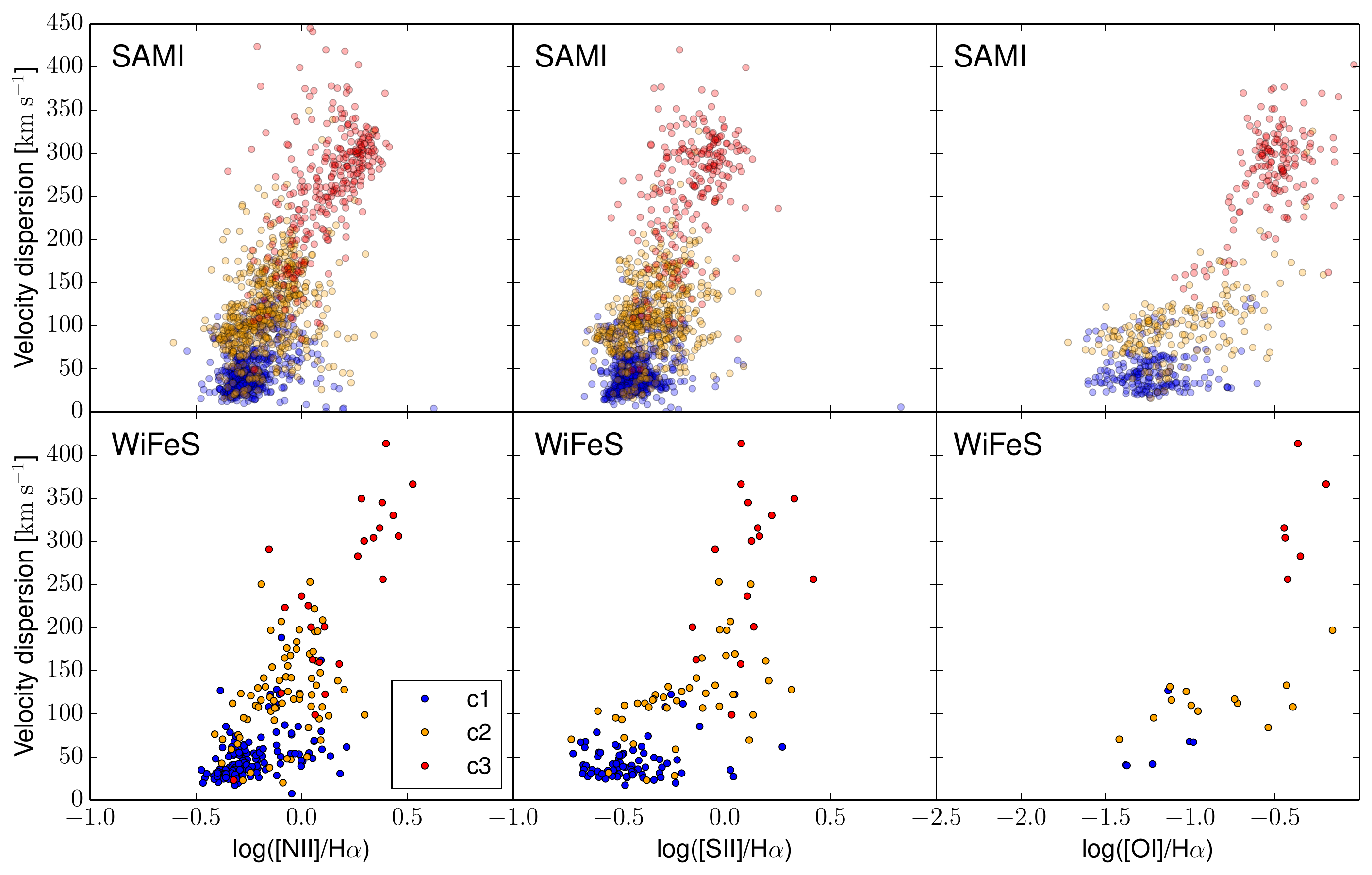}
\caption{Velocity dispersions versus the three key line ratios for the SAMI data (top) and the WiFeS data (bottom). Measurements from the three different components are colour-coded as indicated in the legend. }\label{bpt_vdisp}
\end{figure*}

The ability to robustly separate different kinematic components with spectral decomposition opens up opportunities to investigate the underlying physical mechanisms responsible for exciting different kinematic components.  Different physical mechanisms such as thermal excitation by young stars in HII regions, and non-thermal excitation by AGNs, LINERs, asymptotic giant branch stars or interstellar shocks, can produce distinctly different optical line ratios. The most well-understood tool for investigating different excitations is with the BPT diagrams.

In Fig.~\ref{bpt}, we present the 3 BPT diagrams, \OIII/H$\beta$ versus (1) \NII/H$\alpha$, (2) \SII/H$\alpha$ and (3) \OI/H$\alpha$ for both the SAMI data and the WiFeS data. Only spaxels with $\rm S/N >3 $ on all lines associated with the line ratios are shown. 

We also show various empirical and theoretical lines dividing galaxies excited by different mechanisms. The empirical separation lines between starburst galaxies and AGN-hosted galaxies derived from SDSS are shown as dashed lines \citep{Kauffmann:2003vn}; the theoretical extreme starburst lines calculated from photoionisation and radiation transfer models are shown as solid lines \citep{Kewley:2001lr}; the SDSS empirical separation lines between Seyfert~II and LINER galaxies are shown as dotted lines \citep{Kewley:2006lr}.

Figure~\ref{bpt} demonstrates one of our major results: {\it not only do the different kinematic components have different velocity dispersions, they also present distinctly different line ratios on all three BPT diagrams}. Firstly, the narrow kinematic component {\it c1} mostly lies below the maximum starburst line characterized by \citet{Kewley:2001lr}; 99\%, 99\%, and 88\% of the SAMI spaxels on \NII/H$\alpha$, \SII/H$\alpha$ and \OI/H$\alpha$, respectively, lie in the parameter space where thermal excitation is important. Secondly, the broad kinematic component {\it c3} lies in the parameter space where a non-thermal component must exist to explain the line ratios. The line ratios of \SII/H$\alpha$ and \OI/H$\alpha$ are consistent with LINERs. Finally, the intermediate component, {\it c2}, lies in the transition region between {\it c1} and {\it c3}. 

In Fig.~\ref{bpt_vdisp}, we compare the three BPT line ratios to their corresponding velocity dispersions for both the SAMI and WiFeS data. Again, only spaxels with $\rm S/N > 3$ on all the lines associated with each diagram are shown. Comparing Fig.~\ref{bpt_vdisp} to Fig.~\ref{bpt}, there are considerably more data points because the S/N criteria required for \OIII and the weak H$\beta$ line in Fig.~\ref{bpt} are not applied in Fig.~\ref{bpt_vdisp}. 

Figure~\ref{bpt_vdisp} clearly demonstrates that the three components are well separated both in velocity dispersion space and in line ratio space. Clear correlations between the line ratios and velocity dispersions exist not only between different components, but within the same component. For example, the narrow component {\it c1}, {\it on average}, has both lower line ratios and velocity dispersions than the broad component {\it c3}; and within {\it c3}, those with lower velocity dispersions have lower line ratios than those with high velocity dispersions. 

Such clear separation of different kinematic components on BPT diagrams, and the obvious correlations between the line ratios and velocity dispersions have also been observed in late-stage mergers with galactic-wide shock activities \citep{Rich:2011kx}. 

Comparing the SAMI and the WiFeS results in Fig.~\ref{bpt} and \ref{bpt_vdisp}, we see excellent agreement between the general trends on the diagrams, despite the differences in S/N, spectral resolution, spectral coverage, and spatial resolution between the two datasets. We note that there are considerably fewer data points (spaxels) in the WiFeS results. This is due to both the larger spaxel size and the lower S/N of the WiFeS data. The equivalent noise level of the WiFeS data is approximately $2\ \mbox{--}\ 4$ times higher than the SAMI data. 

\subsection{Electron density}

\begin{figure}
\centering
\includegraphics[width = 8.5cm]{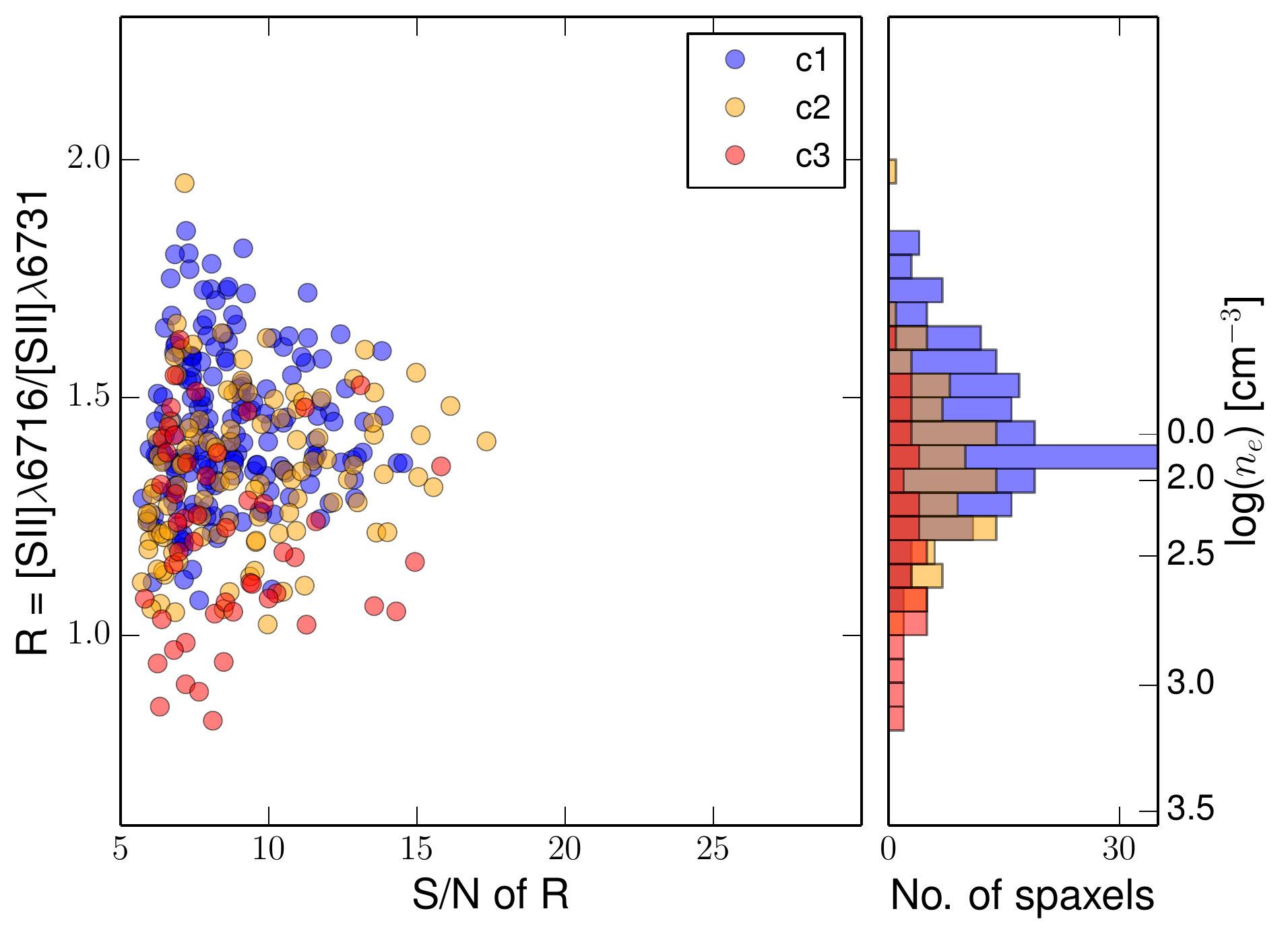}
\caption{{\it Left:} {[\ion{S}{ii}]~$\lambda$6716} to {[\ion{S}{ii}]~$\lambda$6731} line ratios versus S/Ns of the line ratios as measured from the SAMI data. A minimum flux S/N of 8 is applied on both lines. Measurements from different components are colour-coded as indicated in the legend. {\it Right:} Histogram of there data in the left panel. We show the inferred electron densities from the \SII\ ratios on the y-axis on the right-hand-side. Although the density measurements are low in S/N, it is evident that {\it c3} has systematically low electron densities than {\it c1}. }\label{sami_density}
\end{figure}

The density-sensitive {[\ion{S}{ii}]~$\lambda$6716} and {[\ion{S}{ii}]~$\lambda$6731} lines can be used to probe electron density, $n_e$, of the different components useful for shedding light on the different physical environments of the gas. We adopt the {\scshape chianti} package written in {\scshape idl} \citep{Dere:1997lr,Landi:2013fk} to convert the flux ratios {[\ion{S}{ii}]~$\lambda$6716}/{[\ion{S}{ii}]~$\lambda$6731} to electron densities assuming an electron temperature, $T_e$, of $10^4$~K. The distributions of the line ratios and $n_e$ derived from the SAMI data are presented in Fig.~\ref{sami_density}. Here, we apply a harder S/N cut of 8 on both \SII\ lines. The density measurements require high S/Ns on both lines because of the strong dependency of the \SII\ ratio on electron density, and the limited range of \SII\ ratio that returns meaningful electron density. Nevertheless, it is clear that the broad kinematic component {\it c3} has systematically lower line ratios than the narrow component {\it c1}, which infers higher electron densities of {\it c3} than {\it c1}. The median electron density of the broad kinematic component {\it c3} and the narrow kinematic component {\it c1} are approximately 300 and 20 $\rm cm^{-3}$, respectively.

\subsection{Global star formation rates}\label{sec-sfr}

Global star formation rates (SFRs) provide insight into the available energy budget for exciting the emission lines. Global SFRs are estimated from five indicators: (1) H$\alpha$, (2) ultraviolet ({\it UV}) continuum, (3) inferred infrared ({\it IR}) continuum, (4) \CII, and (5) 1.4~GHz radio continuum. 

\subsubsection{H$\alpha$}\label{sec-sfr-halpha}
The H$\alpha$-based SFRs are derived from the narrow component {\it c1}, as {\it c1} originates from star forming regions on the disk (see Section~\ref{sec-c1}). We first estimate the global H$\alpha$ and H$\beta$ fluxes by summing the pixel-to-pixel measurements within the SAMI aperture. We then correct for extinction assuming H$\alpha$/H$\beta$ of 2.85 using the classical extinction law with $R_V$ = 3.1 \citep*[i.e., CCM extinction;][]{Cardelli:1989qy} implemented in the Python module {\scshape pyneb} \citep*{Luridiana:2012ve}. The mean colour excess {\it E(B-V)} is approximately 0.3. Following \citet{Kennicutt:1998zr}, the extinction corrected H$\alpha$ flux infer a SFR of $\rm 1.4~M_\odot~yr^{-1}$ within the SAMI aperture. A higher SFR of approximately $\rm 2.7~M_\odot~yr^{-1}$ is arrived, if we include intermediate component {\it c2} that could also partially originates from star forming regions (see Section~6.3).

\subsubsection{Ultraviolet continuum}
The {\it UV}-based SFR is derived from the far-{\it UV} ({\it FUV}) and near-{\it UV} ({\it NUV}) photometry measured in the {\it Galaxy Evolution Explorer mission} \citep[{\it GALEX};][]{Martin:2005ul,Wyder:2005pd}, which were subsequently compiled by \citet{Hill:2011qf} as part of the GAMA project. Milky-Way extinction is corrected using the reddening map by \citet*{Schlegel:1998ys} and the CCM extinction law \citep{Cardelli:1989qy}. The corrected {\it FUV} and {\it NUV} brightness measure 18.9 and 18.3 magnitudes in the AB system. A dust-corrected {\it UV}-based  SFR of $\rm 5.5~M_\odot~yr^{-1}$ is inferred following the calibration by \citet{Salim:2007ly}. The calibration corrects for internal dust extinction using the {\it UV} colour. 

\begin{table}
\caption{Star Formation Rate}\label{sfr}
\begin{tabular}{cccccc}
\hline
\hline
                & Optical & {\it UV} & $L_{IR}$ & [\ion{C}{ii}]& 1.4~GHz\\
 Indicator&H$\alpha$ & Continuum& & $\lambda158~\mu \rm m$ & Continuum  \\
\hline
SFR& 1.4\ \mbox{--}\ 2.7 & $5.5$ & $6\ \mbox{--}\ 7$ & $15\ \mbox{--}\ 18$ &  $28$ \\
\hline
\end{tabular}
\caption{Star formation rates, in units of $\rm M_\odot~yr^{-1}$, using different indicators. See Section~\ref{sec-sfr} for details.}
\end{table}

\subsubsection{Inferred infrared continuum}

Star formation rate can also be inferred from dust emission in the infrared. The $8\ \mbox{--}\ 1000~\mu\rm m$ luminosity ($L_{IR}$) is commonly calculated from the 4-band photometry measured with the {\it Infrared Astronomical Satellite} \citep[{\it IRAS};][]{Neugebauer:1984cr}; unfortunately those measurements are not available for \shortsource. Since the galaxy is not in the {\it IRAS} faint source catalog \citep*{Moshir:1992wd}, rough upper limits of 0.2~Jy in the 12, 25, 60~$\mu\rm m$ bands, and 1~Jy in the 100~$\mu\rm m$ band imply an upper limit on SFR of $\rm 30~M_\odot~yr^{-1}$ \citep{sanders96,Kennicutt:1998zr}.

A tighter constraint can be placed by comparing templates of spectral energy distribution (SED) with measurements at shorter wavelengths by the {\it Wide-field Infrared Survey Explorer} \citep[{\it WISE};][]{Wright:2010rr}. Comparing the 105 template SEDs constructed by \citet{Chary:2001uq} to the {\it WISE} photometry at 12 and 22$\mu\rm m$, we constrain $L_{IR}$ and derive the corresponding SFR of approximately $\rm 7~M_\odot~yr^{-1}$. If we abandon the $12~\mu\rm m$ measurement that could contain emission from the polycyclic aromatic hydrocarbons, we arrive a similar result of approximately $\rm 6~M_\odot~yr^{-1}$. 

\subsubsection{\CII}
The strong interstellar cooling line \CII\ is an alternative probe of star formation activities. The \CII\ line strength in \shortsource\ is measured with the Photodetector Array Camera and Spectrometer \citep[PACS;][]{Poglitsch:2010vn} aboard the {\it Herschel Space Observatory} (\citealt{Pilbratt:2010eu}; ObsID: 1342270755). Following the calibrations by \citet{de-Looze:2011fj} and  \citet{Sargsyan:2012kx}, the total \CII\ line flux infers SFRs of 15 and $\rm 18~M_\odot~yr^{-1}$, respectively.

\subsubsection{1.4~GHz radio continuum}

Finally, radio continuum originated from synchrotron radiation (from relativistic electrons) and free-free emission from HII regions can also infer SFR. The tight radio to far {\it IR} correlation (\citealt*{Helou:1985fk}; \citealt{Condon:1992qy}; \citealt*{Yun:2001uq}) allows a conversion between the 1.4~GHz radio continuum luminosity and the far {\it IR} luminosity. The infrared luminosity can then be computed by comparing the inferred far {\it IR} luminosity with SED templates, and $L_{IR}$ subsequently yields SFR. Following \citet[][section~3.3.1]{Ho:2010fk} , we convert the 1.4~GHz radio continuum of 5.3~mJy, measured by the Very Large Array FIRST survey \citep{Becker:1994lr}, to a SFR of approximately $\rm28~M_\odot~yr^{-1}$. \\

To summarize, the SFRs derived from various indicators, as tabulated in Table~\ref{sfr}, do not agree well. This is not surprising given that the systematic uncertainties involved in different SFR calibrations \citep[e.g.,][]{Hopkins:2003ly}, and the fact that the different indicators trace star-formation of different timescales; however, an order of magnitude difference is rather extreme. In addition to systematic errors associated with the SFR calibrations, the SFRs from the {\it UV} continuum, 1.4~GHz radio continuum, and \CII emission could be over-estimates because of contamination from shock excitation (see below). The {\it UV} continuum and the H$\alpha$ are subject to uncertainties in the extinction correction. Our infrared luminosity and 1.4~GHz radio continuum could both be affected by the putative AGN and the unknown dust temperature. Given these caveats, our best estimate of the SFR is approximately $\rm 5\ \mbox{--}\ 15~M_\odot~yr^{-1}$.

\section{Nature of the excitation sources}\label{sec-source-nature}

The distinctly different velocity dispersions, velocity fields, and line ratios strongly indicate that the different kinematic components trace gas of  different origins in the galaxy. Their continuity in parameter space, in particular velocity dispersions and line ratios; however, implies possible connections between different components. Below, we discuss the different components in this order: {\it c1}, {\it c3} and {\it c2}. 

\subsection{c1 -- the star-forming component}\label{sec-c1}

All evidence suggests that the narrow component {\it c1} arises from HII regions within the disk. The distribution of the velocity dispersion of {\it c1} peaks at $\rm \sim40~km~s^{-1}$ (Fig.~\ref{sami_vdisp}), in agreement with those observed in normal star-forming galaxies in the SAMI Galaxy Survey. The emission-line ratios of the {\it c1} component are consistent with star-forming regions; $\gtrsim90\%$ of data points (spaxels) lie below the \citet{Kewley:2001lr} maximum starburst line. Thermal excitation from young stars is expected to dominate, but some contributions from non-thermal excitation cannot be ruled out. The clear rotation map (Fig.~\ref{sami_v}) supports our conclusion that the narrow component {\it c1} is from the disk. The regularity of the rotation map further suggests that the galaxy is a kinematically-isolated system without a recent major merger, which is consistent with the lack of tidal features, distorted morphology, or companions on the optical images (Fig.~\ref{pointings}).

\begin{figure}
\centering
\includegraphics[width = 8.5cm]{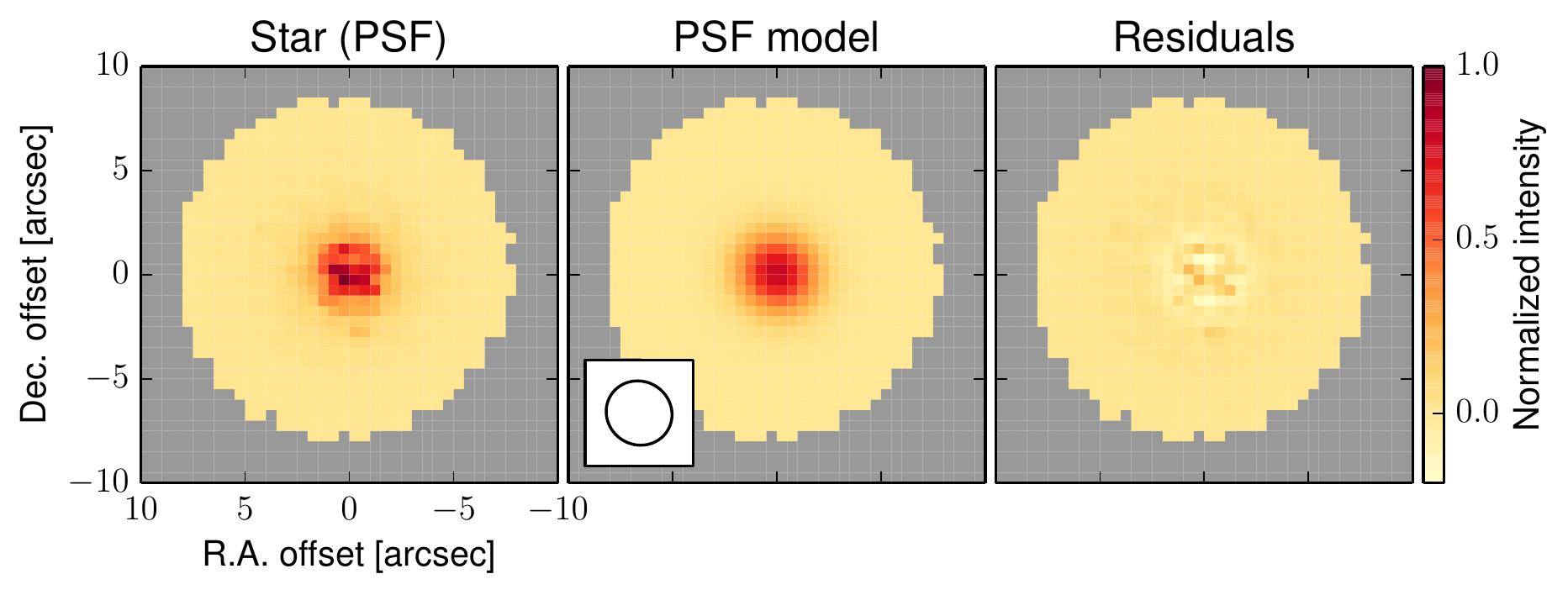}
\caption{{\it Left:} Normalized narrow band image (6400 -- 6500\AA) reconstructed from the data cube of a calibration star. The star was observed simultaneously with the galaxy by SAMI. 
{\it Middle:} Two-dimensional Gaussian fit of the narrow band image. The Gaussian has a FWHM size of $3.2\arcsec\times3.0\arcsec$ and a position angle of $-33^\circ$, as also indicated by the ellipse in the bottom left corner. 
{\it Right:} Residuals of the model fit. 
}\label{sami_psf}
\end{figure}

\begin{figure}
\centering
\includegraphics[width = 8.5cm]{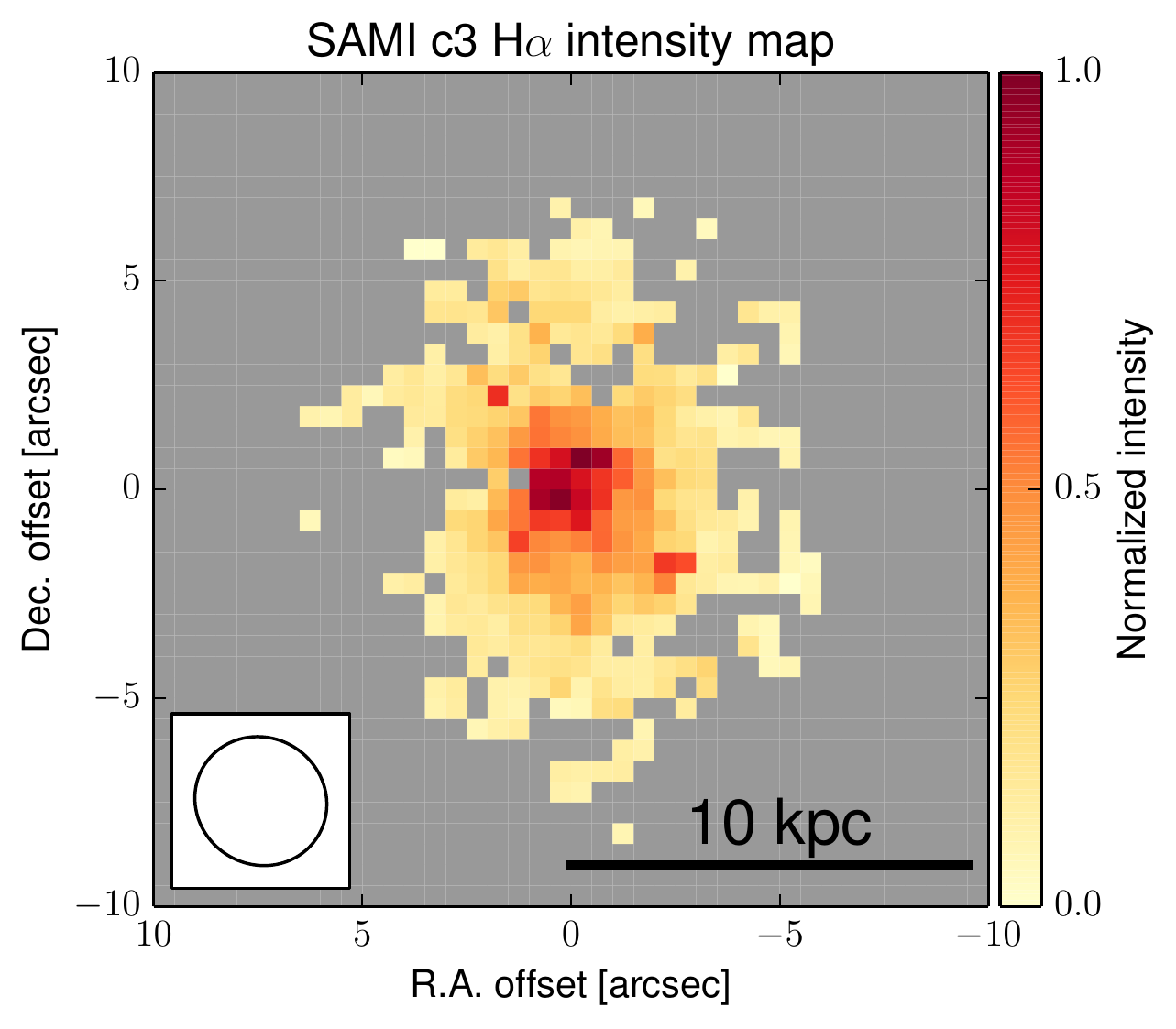}
\caption{Normalized H$\alpha$ intensity map of the broad kinematic component {\it c3}. Spaxels with the H$\alpha$ S/N smaller than 3 have been masked out. The ellipse in the bottom left corner indicates the FWHM resolution as determined from Fig.~\ref{sami_psf}. The broad kinematic component {\it c3} is extended and spatially resolved by SAMI.}\label{sami_c3_halpha}
\end{figure}

\subsection{c3 -- the shock component }\label{sec-c3}

The broad component {\it c3} is markedly different from  {\it c1} in terms of velocity dispersion, velocity, and emission-line ratios. Kinematically, the distribution of velocity dispersions of {\it c3} peaks at a much higher value of $\rm \sim300~km~s^{-1}$ (compared to {\it c1} of $\rm \sim40~km~s^{-1}$; Fig.~\ref{sami_vdisp}), and the velocity gradient of {\it c3} is roughly orthogonal to that of {\it c1} (Fig.~\ref{sami_v}). The optical line ratios of {\it c3} fall in the parameter space classified as LINERs. At the given observational sensitivity, {\it c3} is spatially less extended than {\it c1}, implying the underlying ionisation sources could be located close to the centre of \shortsource.

We believe that {\it c3} is unlikely to arise from the broad line region of a putative Type-I AGN, because the broad line region would yield much broader velocity dispersions ($\rm>1000~km~s^{-1}$; \citealt{Osterbrock:1986lr}; \citealt*{Sulentic:2000fk}). The spatial extent of {\it c3} is also inconsistent with a compact unresolved source. In Fig.~\ref{sami_psf} we show the PSF measured from a calibration star observed simultaneously with the galaxy, and Fig.~\ref{sami_c3_halpha} gives the comparison between the PSF and the H$\alpha$ flux map of {\it c3} from SAMI. The Gaussian-like PSF has a FWHM of approximately $3.1\arcsec$, but the broad {\it c3} kinematic region is non-Gaussian and significantly extended. To quantify the size of the {\it c3} region, we fit a 2D Gaussian to the H$\alpha$ flux map, which yields a FWHM size of approximately $6\arcsec\times4\arcsec$. These dimensions correspond to a deconvolved size of approximately $\rm5.5\times2.8~kpc$.  A consistent result is also obtained with the WiFeS data, which have a better spatial resolution (1\arcsec \mbox{--} 2\arcsec), but approximately $2\ \mbox{--}\ 4$ times  higher noise level. 

The resolved velocity field of {\it c3} shows a velocity gradient almost orthogonal to that of {\it c1}, indicating that {\it c3} is kinematically decoupled from the stellar disk (Fig.~\ref{sami_v}). The high velocity dispersions and the clear velocity gradient of {\it c3} are most naturally explained by bipolar (biconical) outflows seen in many galaxies hosting AGNs and/or star-bursting \citep{Veilleux:2005qy}. If {\it c3} genuinely traces the outflowing gas, the emission lines are likely to be excited by radiative shocks \citep[e.g.,][]{Rich:2010yq,Rich:2011kx,Soto:2012fk,Soto:2012kx}.

\subsubsection{Shock modeling and mixing}\label{sec-mappings}

Radiative shocks can be modeled using {\scshape mappings}, a shock/photoionisation code originally described in \citet{Sutherland:1993lr}. We use the latest {\scshape mappings~iv} code \citep{Dopita:2013qy}, which incorporates a new non-thermal electron energy excitation (\citealt*{Nicholls:2012fk}; \citealt{Nicholls:2013uq}) and updated atomic data, to model the line ratios. The development of the models and efforts of modeling optical line ratios, in particular those on the BPT diagrams, can be found in \citet{Dopita:1995fj,Dopita:1996uq} and \citet{Allen:2008fk}.

We first constrain the chemical abundance of the gas by measuring the chemical abundance of the star-forming gas (i.e., the narrow component {\it c1}). The line fluxes are extinction corrected using the CCM extinction curve \citep{Cardelli:1989qy}. We adopt line ratio diagnostics constructed consistently from starburst photoionisation models in {\scshape mappings~iv} to measure the chemical abundance. The total measured oxygen abundance 12 + log(O/H) is approximately 9.1 -- 9.2, or 2.6 -- 3.2 times the solar abundance, and the ionisation parameter log(q) is approximately 7.3 -- 7.5 $\rm cm~s^{-1}$. The total oxygen abundance 12 + log(O/H) in the shock models is therefore fixed to 9.14. The one-dimensional plane-parallel shock models have a pre-shock hydrogen number density of $\rm 10~cm^{-3}$ and a transverse magnetic field of $10\mu\rm G$. In total, ten shock models are constructed and the only variable is the shock velocity, $v_{sh}$, with increments of $\rm20~km~s^{-1}$ between $\rm100~km~s^{-1}$ and $\rm200~km~s^{-1}$, and increments of $\rm25~km~s^{-1}$ between $\rm200~km~s^{-1}$ and $\rm300~km~s^{-1}$. As a technical note, these models do not allow for grain destruction in the shock, which would increase the gas-phase oxygen abundance by about 0.08~dex.

\begin{figure*}
\centering
\includegraphics[width = 16cm]{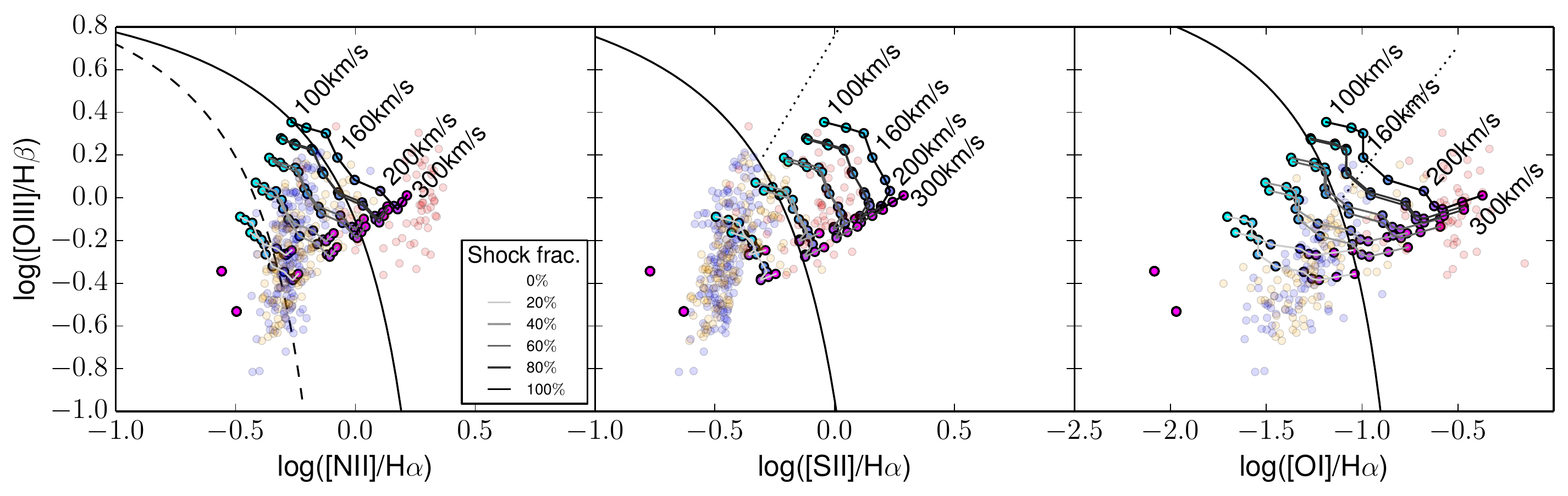}
\caption{Shock and photoionisation mixing sequences predicted by our models. The line segments connect model grids of certain shock fractions with changing shock velocities. Different grey scales of the segments correspond to different shock fractions as indicated in the legend. The shock velocity grids have increments of $\rm20~km~s^{-1}$ between $\rm100~km~s^{-1}$ and $\rm200~km~s^{-1}$, and increments of $\rm25~km~s^{-1}$ between $\rm200~km~s^{-1}$ and $\rm300~km~s^{-1}$. The two sets of 0\% shock fraction points (lower left corners) are models of pure photoionisation of HII regions of log(q) = 7.5 ($\rm cm~s^{-1}$; upper data point) and 7.25 ($\rm cm~s^{-1}$; lower data point), and 12+log(O/H) = 9.14 \citep{Dopita:2013qy}. Blue, orange, and red transparent data points are SAMI measurements of the different kinematic components shown in Fig.~\ref{bpt}. Details of the models are provided in Section~\ref{sec-mappings}. }\label{bpt_model}
\end{figure*}

We combine the {\scshape mappings~iv} shock and photoionisation models to account for the potential contribution of both shocks and star-formation to individual spaxels. Here, we define a shock fraction, $f_{sh}$, representing the contribution of shock excitation relative to HII regions. We mix the model line fluxes by 
\begin{equation}\label{eq-bpt-mixing}
F_{mix}(f_{sh}) = F_{sh} f_{sh} + F_{HII} (1-f_{sh}), 
\end{equation}
where $F_{sh}$ and $F_{HII}$ are the line fluxes from the {\scshape mappings~iv} models, and  $F_{mix}(f_{sh})$ is the  (mixed) flux contributed by both shock and photoionisation. All the line fluxes are normalized to H$\beta$. The photoionisation models have an oxygen abundance, 12+log(O/H), of 9.14 and two possible ionisation parameters, log($q$), of 7.25 and 7.5 ($\rm cm~s^{-1}$). In Fig.~\ref{bpt_model}, we show our shock and photoinoization mixing models. We discuss the physical meanings in more detail later.

We note that the line ratios of pure photoionisation ($f_{sh} = 0$) are generally in agreement within approximately 0.3~dex with the star-forming component {\it c1} of the lowest \OIII/H$\beta$ values. The only exception is \OI/H$\alpha$ where the model is approximately 0.6~dex smaller than the observations. \citet{Dopita:2013qy} state that the \OI\ models are less reliable than the other two line ratios because \OI\ is sensitive to even very small mechanical energy injection (see their section~7.1). From our data, it is not clear whether this difference is due to numerical artefacts or contaminations from minor shock activities. Our conclusions do not change even if we abandon the theoretical \OI/H$\alpha$ and adopt empirical values draw directly from the star-forming component {\it c1} where the shock contribution is expected to be minimal.

An important quantity measured observationally but not predicted in the shock models is velocity dispersion. While the measured velocity dispersion may depend on shock geometry, a positive correlation between the measured velocity dispersion and the shock velocity is expected when there are multiple shocks propagating in virtually random directions in each spaxel. Given that our data have insufficient spatial resolution to resolve the shock fronts, we expect a close relationship between the measured velocity dispersion and the shock velocity.   A closer examination of this possibility can be found in (\citealt{Dopita:2012fr}; their section~3.2) where indeed a positive correlation was discovered. 

\begin{figure*}
\centering
\includegraphics[width = 16cm]{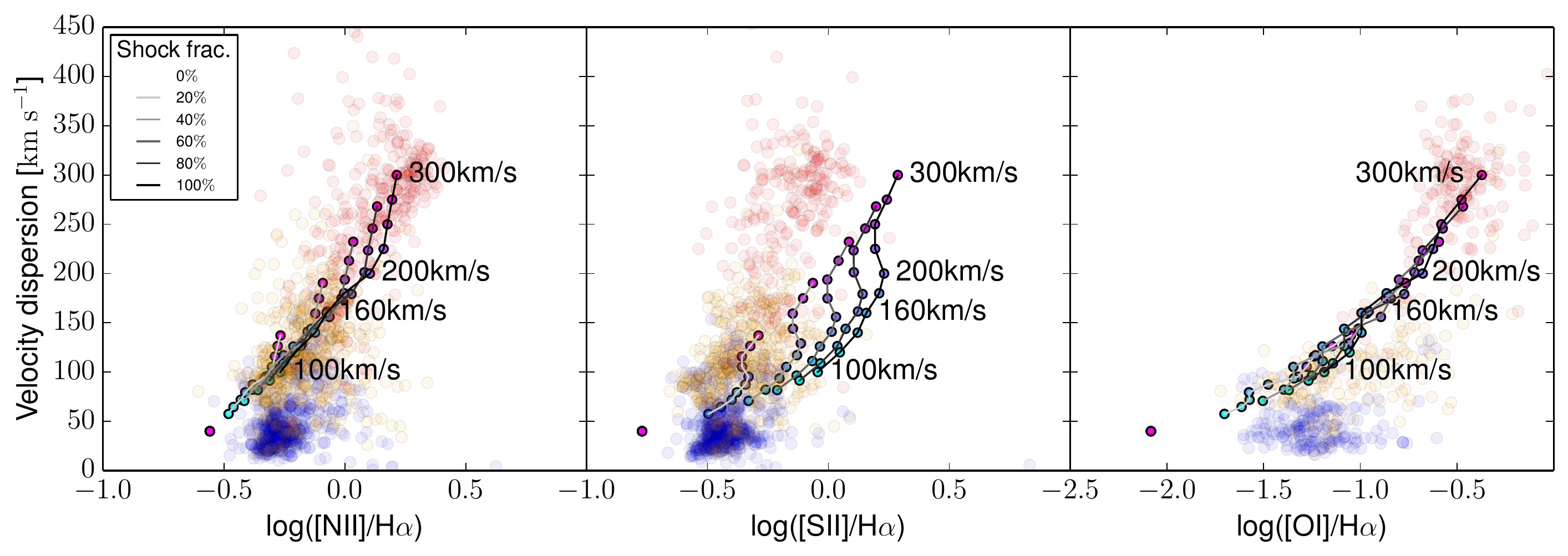}
\caption{
Velocity dispersion mixing sequence as predicted by our photoionisation and shock models. The line segments connect model grids of certain shock fractions with changing shock velocities. Greyscales of the segments correspond to different shock fractions as indicated in the legend. The shock velocity grids have increments of $\rm20~km~s^{-1}$ between $\rm100~km~s^{-1}$ and $\rm200~km~s^{-1}$, and increments of $\rm25~km~s^{-1}$ between $\rm200~km~s^{-1}$ and $\rm300~km~s^{-1}$. The 0\% shock fraction points (lower left corners) are models of pure photoionisation of HII regions of log(q) = 7.5 $\rm cm~s^{-1}$, and 12+log(O/H) = 9.14 \citep{Dopita:2013qy}. Blue, orange, and red transparent data points are SAMI measurements of the different kinematic components shown in Fig.~\ref{bpt}. Details of the models are provided in Section~\ref{sec-mappings}.}\label{bpt_vdisp_model}
\end{figure*}

In consideration of this correlation, we {\it assume} that the model shock velocities equal to the observed velocity dispersions. Under this assumption, we can predict the velocity dispersion mixing sequences in Fig.~\ref{bpt_vdisp}, very much like the mixing sequences on BPT diagrams in Fig.~\ref{bpt} using Equation~\ref{eq-bpt-mixing}. We define the mixing of velocity dispersions between the shock and photoionisation models,  $\sigma_{mix}(f_{sh})$, as 
\begin{equation}\label{eq-bpt-vdisp-mixing}
\sqrt{\frac{v_{sh}^2 F_{H\alpha,sh} f_{sh} + \sigma_{HII}^2 F_{H\alpha, HII}(1 - f_{sh})} {F_{H\alpha,sh} f_{sh} + F_{H\alpha, HII}(1 - f_{sh})} } , 
\end{equation}
where $F_{H\alpha,sh}$ and $F_{H\alpha,HII}$ are the normalized line fluxes of H$\alpha$ from {\scshape mappings~iv}, and $\sigma_{HII}$ is the velocity dispersion of the star-forming component {\it c1} assumed at its peak value of $\rm 40~km~s^{-1}$ (Fig.~\ref{sami_vdisp}). The velocity dispersion mixing sequences, $\sigma_{mix}(f_{sh})$, from the photoionisation  and shock models are shown in Fig.~\ref{bpt_vdisp_model}. \\

The line ratios of {\it c3} are readily explained by the shock models. In Fig.~\ref{bpt_model}, the high \NII/H$\alpha$, \SII/H$\alpha$, and \OI/H$\alpha$ spaxels are consistent with  being excited only by higher velocity shocks ($v_{sh} \approx \rm 200\ \mbox{--}\ 300~km~s^{-1}$). The spaxels with relatively low line ratios can be explained by being excited by shocks of lower velocities and slight mixing with HII regions ($f_{sh}\approx 60\ \mbox{--}\ 80\%$). 
Spaxels from all three kinematic components can be explained by the models; the model grids encompass 60\%, 70\%, and 55\% (for \NII/H$\alpha$, \SII/H$\alpha$, and \OI/H$\alpha$, respectively) of all the SAMI data point in Fig.~\ref{bpt_model}. 

Remarkably, the positive correlations between the three key BPT line ratios and velocity dispersions are also qualitatively reproduced by the model loci in Fig.~\ref{bpt_vdisp_model}. These results demonstrate that our models simultaneously reproduce both the kinematics and the line ratios, with the shock velocity as the only variable.

\begin{figure*}
\centering
\includegraphics[width = 16cm]{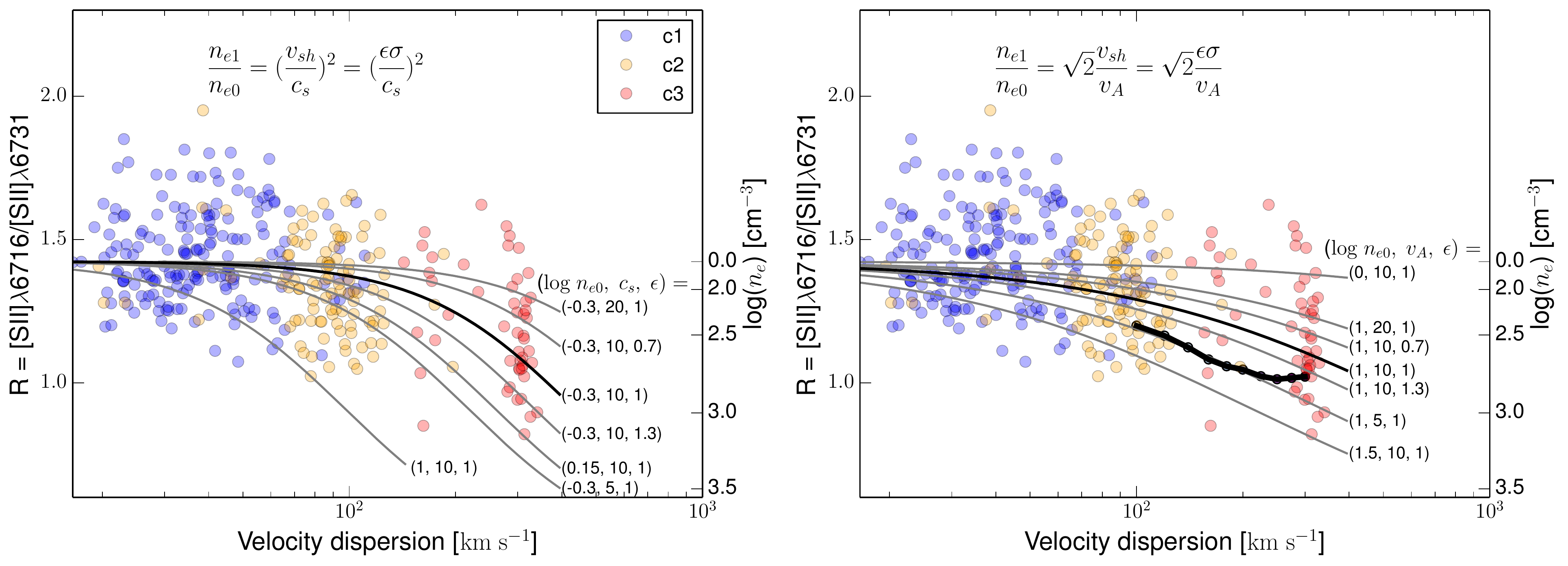}
\caption{\SII\ ratios and electron densities, versus velocity dispersion for the SAMI data shown in Fig.~\ref{sami_density}. Filled circles of different colours represent different components as shown in the legend. The curves are models of isothermal shocks propagating in ISM without magnetic field (left panel) and with magnetic field (right panel). The models described by Equation~\ref{eq-shock} and \ref{eq-b-shock} are also labeled in the two panels. The models providing the best representation of the data are shown as the black curves. The models with different parameters close to those of the black models are shown in grey. The three parameters adopted in each model are labeled next to each curve. The units are $\rm cm^{-3}$ for the pre-shock electron density ($n_{e0}$), and $\rm km~s^{-1}$ for the sound speed ($c_s$) and Alfv\'{e}n velocity ($v_A$). The geometric factor ($\epsilon$) is unit-less. The predictions from the {\scshape mappings~iv} shock models are shown as thick black segments in the right panel. Details are discussed in Section~\ref{sec-mappings}. }\label{sami_density_vdisp}
\end{figure*}

The high electron densities of the {\it c3} region also strongly support the presence of shocks. The density indicators [\ion{S}{ii}]~$\lambda\lambda$6716,31 come from the post-shock recombination regions where the gas densities are expected to be higher than the pre-shock densities \citep[e.g., figure~12 in][]{Allen:2008fk}. In the simplest case, where the shocks are isothermal and travel in non-magnetized medium, the density contrast between pre-shock and post-shock regions follows
\begin{equation}\label{eq-shock}
\frac{n_{e1}}{n_{e0}} = (\frac{v_{sh}}{c_s})^2 = (\frac{\epsilon \sigma}{c_s})^2, 
\end{equation}
This is sometimes also called the `compression factor'. Here, $n_{e0}$ and $n_{e1}$ are pre- and post-shock electron densities, respectively. The sound speed is represented as $c_s$. We denote $v_{sh}$ as the product of a geometric factor, $\epsilon$, and the observed velocity dispersion, $\sigma$. The geometric factor represents a zeroth order approximation for the relation between shock velocities and velocity dispersions which we adopted as unity when we calculated the mixing sequences. 

In the case where the transverse magnetic field pressure dominates the gas pressure in the recombination zone of the shock, the density contrast becomes
\begin{equation}\label{eq-b-shock}
\frac{n_{e1}}{n_{e0}} = \sqrt{2} \frac{v_{sh}}{v_{A}}  = \sqrt{2} \frac{\epsilon\sigma}{v_{A}}, 
\end{equation}
where $v_{A}$ is the Alfv\'{e}n velocity determined by the density and magnetic field \citep[see][]{Dopita:2003fr}.  

In Fig.~\ref{sami_density_vdisp}, we present the \SII\ line ratios and electron densities versus velocity dispersions from the SAMI data. We also show the simple models from Equation~\ref{eq-shock} (left panel) and Equation~\ref{eq-b-shock} (right panel). Models of different parameters are shown as solid (grey) curves. The models providing the best representation of the data are shown as the black curves. In the case of a shock propagating in a medium without magnetic field (left panel), our {\it c3} component can be explained by compression from few hundred $\rm km~s^{-1}$ shocks propagating in a medium of density of $\rm 10^{-0.3}~cm^{-3}$ and sound speed of $\rm10~km~s^{-1}$. Such a small density, although not impossible, is at the lower boundary of typical ISM conditions. In the case where magnetic field is important, the medium can resist compression by virtue of the magnetic pressure and therefore the pre-shock density can be higher. Our preferred model (thick black curve in the right panel) has a pre-shock density of $\rm 10~cm^{-3}$ and an Alfv\'{e}n velocity of $\rm 10~km~s^{-1}$, which corresponds to a pre-shock (transverse) magnetic field of approximately $18\mu\rm G$. In this case, the pre-shock density is more consistent with typical ISM values, and both the density and (transverse) magnetic field agree with the values adopted in the shock models of $\rm 10~cm^{-3}$ and $\rm 10\mu\rm G$, respectively. In fact, the {\scshape mappings~iv} predictions shown in the right panel of Fig.~\ref{sami_density_vdisp} are also largely consistent with the data.

The fact that the {\scshape mappings~iv} models are able to simultaneously reproduce (1) the line ratios on BPT diagrams, (2) the velocity dispersion versus line ratio relations, and (3) the high densities of {\it c3} is remarkable, given that all parameters are either observationally or empirically constrained and the {\it only} free parameter is the shock velocity. It is clear that shock model velocities can be inferred from the observed velocity dispersions for the {\it c3} component of this galaxy.

\subsection{c2 -- the star-forming \&  shock mixed component}\label{sec-c2}

The intermediate kinematic component {\it c2} has velocity dispersions, velocity fields, and line ratios that are intermediate between our star-forming component {\it c1} and our shock component {\it c3}.  These intermediate properties suggest that either the {\it c2} component contains contributions from both HII regions and shocks, or that {\it c2} is a transition region between the two excitation sources. 

In Fig.~\ref{bpt_model}, the line ratios of {\it c2} are consistent with excitation by shocks of lower velocities ($v_{sh} \approx 100\ \mbox{--}\ 200~\rm km~s^{-1}$) with lower shock fractions ($f_{sh} \lesssim 60\ \mbox{--} \ 80\%$), compared to the pure shock component ({\it c3}). In Fig.~\ref{bpt_vdisp_model}, the observed correlations between the line ratios and the velocity dispersion can also be explained by the low shock velocities and the low shock fractions, though inferring the velocities and the shock fractions of {\it c2} is more affected by the degeneracy between the two quantities on the line ratio \mbox{--} velocity dispersion space. In Fig.~\ref{sami_density_vdisp}, the lower densities of {\it c2} versus {\it c3} further support the picture that {\it c3} is highly compressed by high velocity shocks and {\it c2} is only moderately compressed due to lower shock velocities. 

We note that some intermediate components could come from star-forming gas, particularly those with virtually zero shock fractions and line ratios consistent with the narrow component {\it c1}. Beam smearing due to poor PSF can sometimes produce skewed line profiles when there is a steep velocity gradient \citep[e.g.,][]{Green:2014ys}. The skewed line profiles would be decomposed into two (narrow and broad) kinematic components with indistinguishable BPT line ratios.

\subsection{Interpretation in terms of a wind model}
What exactly do {\it c2} and {\it c3} trace in the bipolar outflow picture and what are the differences? In the canonical view of bipolar outflows, the outflow velocity in open-ended conical winds often increases with radius. This rise in wind velocity is both expected from a theoretical viewpoint (\citealt*{Murray:2005fk}; \citealt{Murray:2011fj}) and observed in nearby starburst galaxies launching outflows \citep[e.g.,][]{Shopbell:1998yq,Westmoquette:2011uq}. Typical wind velocities of a few hundred $\rm km~s^{-1}$ commonly observed in starburst galaxies is consistent with the inferred shock velocities of {\it c3}. 
This suggests that {\it c3} is at high galactic latitude where the wind-blown bubbles have already broken out of the stellar disk and the outflows are predominately pressure driven. Warm and cool gas entrained by the outflows are compressed by fast shocks propagating in the outflows, which results in the high densities and line ratios of {\it c3}. Because the material is at high galactic latitude, ionising radiation from young stars could be negligible and the excitation would be dominated by shocks. 

Closer to the outflow launching points and at low galactic latitude, the environment is more clumpy and higher density on average; therefore, the outflows do not have sufficient time and space to accelerate. Shock velocities are slower and the interstellar medium experiences both shock excitation from stellar and supernova winds, and photoionisation from nearby young stars and star clusters. Turbulent mixing layers form on the surfaces of the molecular clouds directly facing the winds from young stars and star clusters could also give rise to the intermediate component {\it c2} with shock and star-forming mixed properties (see figure~16 of \citealt{Westmoquette:2009uq}). This picture readily explains the intermediate line ratios, velocity dispersions, and lower shock fractions of {\it c2}. 

We note that in classical nearby wind galaxies surface brightness of the HII emission typically decreases faster with increasing radius than the shock-excited emission \citep[e.g.,][]{Sharp:2010qy}. The fact that the star-forming component {\it c1} is well detected (S/N of H$\alpha > 3 $) within the SAMI aperture but the shock emission {\it c3} is not detected in the eastern and western edges of the aperture (the outflow direction; Fig.~\ref{sami_v}) implies that either \shortsource\ is relatively face-on, and/or the outflows are still developing and did not have sufficient time to expand well beyond the optical disk when viewed in projection on the sky.

\section{Discussion}\label{sec-discussion}

\subsection{Starburst-driven or AGN-driven outflows?} 
Determining the exact energy source driving the outflows, whether from an AGN or starburst, has never been easy. Both energy injections from an AGN or central starburst can yield very similar bipolar structures, although the orientations of AGN-driven outflows can be random relative to the major axis of the host galaxy (e.g., \citealt{Ulvestad:1984uq}; \citealt*{Cecil:1990qy}; \citealt{Kinney:2000fj}). To completely reject the involvement from an AGN is non-trivial given its small physical scale and large dynamic range. Without spatially resolving the base of the outflows, the dominant energy source has to be inferred from indirect evidence. Studies of outflowing neutral gas in starburst and Seyfert galaxies show strong correlations between the outflow detection rate and the far {\it IR} luminosity (\citealt*{Rupke:2005th}; \citealt{rupke:2005b}). At a similar far {\it IR} luminosity, however, \citet*{Krug:2010fj} do not find higher detection rates in Seyfert galaxies than in starburst galaxies, indicating no strong evidence for the AGN dominating large-scale outflow dynamics. The correlation between AGN luminosity and maximum velocity of molecular outflows in nearby ULIRGs, however, implies that AGN could be the powering source for materials of the highest velocities \citep{Sturm:2011fj,Veilleux:2013uq}. 

The nuclear spectrum measured by SDSS (Fig.~\ref{bpt}) does not support the existence of a Type-II AGN, instead implying that \shortsource\ could be a `transition object'. The nuclear line ratios fall within 0.1~dex to the separation line of HII and LINER, indicating that a pure LINER nucleus and HII regions are both presented in the SDSS aperture.  Our results strongly suggest that the LINER-like line ratios arise from shock excitation.  The LINER emission may also arise from non-stellar photoionisation from a low-luminosity AGN (LLAGN) with a low accretion rate \citep{Kewley:2006lr}, which we cannot rule out. If there {\it is} a LLAGN, is it capable of driving the observed large scale galactic outflows by its own?

From energy considerations, we compare the total energy output from the putative LLAGN to the mechanical luminosity of the outflowing gas. The bolometric luminosity of the nuclei in nearby transition objects has a distribution characterized by mean, error and median of $3.0\times10^{40}\rm\ erg~s^{-1}$, $9.7\times10^{39}\rm\ erg~s^{-1}$, $6.5\times10^{39}\rm~erg~s^{-1}$, respectively \citep{Ho:2009qy}. 
To estimate the flux of the mechanical energy through shocks, we correct for extinction for the broad component {\it c3} (as in Section~\ref{sec-sfr-halpha}) and calculate the mechanical luminosity from the H$\alpha$ flux using our {\scshape mappings~iv} shock models. The total H$\alpha$ luminosity of $\rm 4\times10^{40}~erg~s^{-1}$ infers a total mechanical luminosity of approximately $\rm (1.0\ \mbox{--}\ 2.5)\times10^{42}~erg~s^{-1}$, assuming shock velocities of 200 -- 300~$\rm km~s^{-1}$. For comparison, the mechanical luminosities carried by neutral gas at low temperatures, as traced by NaID absorption, were measured on the order of $\rm 10^{41}\ \mbox{--}\ 10^{42}~erg~s^{-1}$ in starburst-dominated LIRGs at $z<0.5$ \citep{rupke:2005b}. The mechanical luminosities carried by gas in other phases in \shortsource\ are unknown, but the mechanical luminosity from shocks already far exceeds the available energy from the putative LLAGN. It is unlikely that the outflows in \shortsource\ are driven by the LLAGN unless, within approximately the dynamical timescale of the outflowing gas ($\rm10^{7}~yr$), \shortsource\ was a Seyfert galaxy where the AGN can be $10^3$ times more energetic than a LLAGN \citep{Ho:2009qy}.

In addition, our broad component {\it c3} has a projected size of approximately $\rm5.5\times2.8~kpc$ (Fig.~\ref{sami_c3_halpha}).  Outflows in narrow line regions (NLRs) in Seyfert galaxies are typically confined to the kiloparsec to sub-kiloparsec scale \citep{Bennert:2006fj,Bennert:2006kx,Fischer:2013yq}, and NLRs in LINERs are smaller, few tens to few hundred parsecs \citep{Masegosa:2011tg}. 

To constrain the mechanical luminosity by star-formation, we adopt the estimate of SFR of approximately $\rm 5\ \mbox{--}\ 15~M_\odot~yr^{-1}$ in Section~\ref{sec-sfr}, and assume the time evolution of the mass-loss rates and mechanical luminosities from {\scshape Starburst99} (\citealt{Leitherer:1999fk}; equation~2 in \citealt{Veilleux:2005qy}). The mechanical luminosity of the stellar ejecta in \shortsource\ is approximately $\rm 3.5\times10^{42}\ \mbox{--}\ 1\times10^{43}~ergs~s^{-1}$, which is enough to provide the mechanical luminosity observed in the shock excited component {\it c3} of approximately $\rm (1.0\ \mbox{--}\ 2.5)\times10^{42}~erg~s^{-1}$. We conclude that star formation alone is sufficient (and likely) to drive the outflows.

\citet{Sharp:2010qy} use the ionization properties of the wind material to identify the likely source of the ionization. They show that starburst and AGNs in $L_\star$ galaxies produce roughly comparable numbers of ionizing photons over the lifetime of the `event'. But the time-dependence of the ionizing radiation is very different. In an impulsive burst, hot and young stars radiate {\it UV} before the supernovae drive the hot winds. By the time the wind filaments emerge, most of the stellar ionizing radiation has disappeared. In contrast, AGN appear to radiate for longer than a typical starburst event, and can non-thermally ionize filaments on much longer timescales. Thus, in their paradigm, starburst-driven winds are likely to be dominated by shocks at large radius. This conclusion is consistent with our finding here. The picture of impulsive starbursts in the past is also consistent with the higher radio- than H$\alpha$-inferred SFR, because the radio emission predominately coming from supernova remnants traces the star-formation in the past ($\rm 10^7\ \mbox{--}\ 10^8~yr$) whereas the H$\alpha$ emission excited by young hot stars traces the current star-formation. 

\citet{Soto:2012fk} and \citet{Soto:2012kx} observed a sample of 39 ULIRGs ($z\sim0.04\ \mbox{--}\ 0.15$) at various merger stages with long-slit spectroscopy. The majority of their sample presents global BPT line ratios consistent with star-forming galaxies (starburst: 43\%; LINER: 18\%; Seyfert: 12\%; and ambiguous: 27\%). They perform a very similar 2-component emission line decomposition, and find that the broad ($\sigma>150\rm~km~s^{-1}$), blueshifted emissions commonly seen in their spectra have 1) emission line ratios consistent with shock excitation ($v_{sh}\approx200\ \mbox{--}\ 300\rm~km~s^{-1}$), and 2) can extend many kiloparsecs ($\rm\gtrsim5~kpc$) from the nucleus, indicating that outflows are very common in merging ULIRGs \citep[see also ][]{Rupke:2013zr}. With these galaxies forming stars at rates of approximately $50\ \mbox{--}\ 200\rm~M_\odot~yr^{-1}$, \citet{Soto:2012kx} argue that the shock energetics predominately come from supernova feedback and stellar winds. It is perhaps intriguing to see that with the much lower current SFR of approximately $\rm 5\ \mbox{--}\ 15~M_\odot~yr^{-1}$, \shortsource\ still presents wind and shock signatures very similar to those seen in the merger sample. In 14 nearby dwarf galaxies of even lower SFRs ($\ll 0.5~\rm M_\odot~yr^{-1}$), \citet{Martin:1998qy} shows that expanding shells of ionised gas are common in the \NII\ and H$\alpha$ narrow band images. In some cases, their long-slit spectra suggest that the diffuse warm ionised gas requires an additional shock component to explain the excitation \citep{Martin:1997yq}. The coming large IFS survey will address the extent to which starburst-driven winds can persist with decreasing SFRs, and what properties in addition to SFR (or SFR surface density) may govern the ability to launch outflows.

\section{Conclusions}\label{sec-conclusions}

We study the nature of \shortsource\ ($z=0.05386$) using IFS data from the SAMI Galaxy Survey, supplemented by IFS data from WiFeS. The SDSS nuclear line ratios indicate a composite/ambiguous galaxy when classified using traditional optical diagnostic diagrams. In IFS data, however, the galaxy presents skewed line profiles changing with position in the galaxy, and emission-line ratios that correlate with the galaxy radius. We show that the skewed line profiles come from different kinematic components overlapping in the line-of-sight direction. We perform spectral decomposition assisted by statistical tests to separate the different kinematic components, and we find that each pixel can either be described by the combination of (1) a narrow kinematic component consistent with HII regions, (2) a broad kinematic component consistent with shock excitation, and (3) an intermediate component consistent with shock excitation and photoionisation mixing. 

We present kinematics, line ratios, and electron densities of different components. We find that

\begin{itemize}
\item[1.] The three components show distinctly different velocity dispersions and  velocity fields. The narrow component traces a regular rotation pattern of the stellar disk, while the broad component presents a velocity gradient almost perpendicular to the disk rotation, consistent with large-scale gas outflows. 
\item[2.] The three kinematic components are clearly separated on the standard optical diagnostic diagrams.  The narrow component is consistent with photoionisation from young hot stars, while the broad component has significantly larger line ratios, consistent with a harder ionising radiation field.
\item[3.] There are positive correlations between velocity dispersions and the three key BPT line ratios (\NII/H$\alpha$, \SII/H$\alpha$, and \OI/H$\alpha$). The correlations exist both between the averages of the different kinematic components, and between different spaxels within the same kinematic component. 
\item[4.] The electron density of the narrow kinematic component (20 $\rm cm^{-3}$) is consistent with HII regions, while the largest electron densities (300 $\rm cm^{-3}$) are found in the broad kinematic component.
\end{itemize}

To interpret the distinctly different line ratios, electron densities and velocity dispersions between different components, we compare our observations with new predictions from the {\scshape mappings~iv} shock and photoionisation models. With all parameters constrained either empirically or observationally and only one free parameter, the shock velocity, the shock models successfully reproduce the line ratios, velocity dispersions, and electron densities. We find that

\begin{itemize}
\item[1.] The broad kinematic component traces regions excited by shocks with velocities of  $\approx200\ \mbox{--}\ 300\rm~km~s^{-1}$. Both the line ratios and velocity dispersions are reproduced by the shock models. We show that the high densities of the broad component are direct results of compression of the interstellar medium by the shock fronts. We conclude that our broad component traces the limb-brightened emission in biconical outflows at high galactic latitude.
\item[2.] The intermediate kinematic component  traces regions excited by low velocity shocks ($\approx100\ \mbox{--}\ 200\rm~km~s^{-1}$), and contains emission from both shocks and HII regions. We believe that the intermediate component traces emission from the base of the outflows at low galactic latitude
\end{itemize}

We argue from the energy considerations that, with the lack of a powerful central AGN, the outflows in \shortsource\ are likely to be driven by starburst activities.

The remarkable agreement between our integral field data and the theoretical shock and photoionisation models sets a benchmark of what can be achieved by the SAMI Galaxy Survey. An important implication from our results is that large integral field spectroscopic surveys, such as the SAMI and MaNGA surveys, are likely to provide deep insight into the prevalence and cause of galactic-scale outflows in the local Universe. 

\section*{Acknowledgments}

We thank the anonymous referee for providing helpful comments. M.A.D. and L.J.K. acknowledge the support of the Australian Research Council (ARC) through Discovery project DP130103925 and DP130104879. This work was also funded in part by the Deanship of Scientific Research (DSR), King Abdulaziz University, under grant No. (5-130/1433 HiCi). L.J.K. gratefully acknowledges the support of an ARC Future Fellowship.  M.S.O. acknowledges the funding support from the Australian Research Council through a Super Science Fellowship (ARC FS110200023). We thank the time allocation committee of the 2.3-m telescope operated by the Australian National University for the support of this work. We also thank Marja Seidel, Ra\'{u}l Cacho, and Fr\'{e}d\'{e}ric Vogt for helping with the WiFeS data reduction. I.-T.H. thanks Mark Westmoquette and Wei-Hao Wang for useful discussions. I.-T.H. also acknowledges the hospitality of the Research School of Astronomy and Astrophysics, Australian National University, during a major portion of this work. I.S.K. is the recipient of a John Stocker Postdoctoral Fellowship from the Science and Industry Endowment Fund (Australia). J.T.A. acknowledges the award of an ARC Super Science Fellowship through project FS110200013. 

This publication makes use of data products from the {\it Wide-field Infrared Survey Explorer}, which is a joint project of the University of California, Los Angeles, and the Jet Propulsion Laboratory/California Institute of Technology, funded by the National Aeronautics and Space Administration. 

The SAMI Galaxy Survey is based on observation made at the Anglo-Australian Telescope. The Sydney-AAO Multi-object Integral field spectrograph was developed jointly by the University of Sydney and the Australian Astronomical Observatory. The SAMI input catalogue is based on data taken from the Sloan Digital Sky Survey, the GAMA Survey and the VST ATLAS Survey. The SAMI Galaxy Survey is funded by the Australian Research Council Centre of Excellence for All-sky Astrophysics, through project number CE110001020, and other participating institutions. The SAMI Galaxy Survey website is \href{http://sami-survey.org/}{http://sami-survey.org/}.

\bibliography{/Users/itho/Dropbox/references}

\end{document}